\documentclass{pasj00}

\def\commenta{$^*$}
\def\commentb{$^\dagger$}
\def\commentc{$^\ddagger$}
\def\commentd{$^\S$}
\def\commente{$^\|$}

\def\submitted{submitted}
\def\inpress{in press}

\def\arxiv#1{ (arXiv astro-ph/#1)}

\DeclareAbbreviation\AAHam{Astron. Abh. Hamburg. Sternw.}
\DeclareAbbreviation\AARv{Astron. Astrophys. Rev.}
\DeclareAbbreviation\AAS{American Astron. Soc. Meeting Abstracts}
\DeclareAbbreviation\AcA{Acta Astron.}
\DeclareAbbreviation\actaa{Acta Astron.}
\DeclareAbbreviation\Afz{Astrofizika}
\DeclareAbbreviation\AGAb{Astronomische Gesellschaft Abstract Ser.}
\DeclareAbbreviation\an{Astron. Nachr.}
\DeclareAbbreviation\AnAp{Annales d'Astrophysique}
\DeclareAbbreviation\AnTok{Tokyo Astron. Obs. Annals, Sec. Ser.}
\DeclareAbbreviation\Ap{Astrophysics}
\DeclareAbbreviation\ARep{Astron. Rep.}
\DeclareAbbreviation\AstBu{Astrophys. Bull.}
\DeclareAbbreviation\ATel{Astron. Telegram}
\DeclareAbbreviation\ATsir{Astron. Tsirk.}
\DeclareAbbreviation\AcApS{Acta Astrophys. Sinica}
\DeclareAbbreviation\AstL{Astron. Lett.}
\DeclareAbbreviation\BaltA{Baltic Astron.}
\DeclareAbbreviation\BANS{Bull. of the Astron. Institutes of the Netherlands Suppl. Ser.}
\DeclareAbbreviation\BASI{Bull. Astron. Soc. India}
\DeclareAbbreviation\BeSN{Be Newslett.}
\DeclareAbbreviation\BHarO{Harvard Coll. Obs. Bull.}
\DeclareAbbreviation\CBET{Cent. Bur. Electron. Telegrams}
\DeclareAbbreviation\ChJAA{Chinese J. of Astron. and Astrophys.}
\DeclareAbbreviation\caa{Chinese J. of Astron. and Astrophys.}
\DeclareAbbreviation\CoAsi{Asiago Contr.}
\DeclareAbbreviation\CoSka{Contributions of the Astronomical Observatory Skalnat\'e Pleso}
\DeclareAbbreviation\GCN{GRB Coord. Netw. Circ.}
\DeclareAbbreviation\ErgAN{Erg. Astron. Nachr.}
\DeclareAbbreviation\ibvs{IBVS}
\DeclareAbbreviation\IEEEP{IEEE Proc.}
\DeclareAbbreviation\JAD{J. Astron. Data}
\DeclareAbbreviation\JApA{J. of Astrophys. and Astron.}
\DeclareAbbreviation\JAVSO{J. American Assoc. Variable Star Obs.}
\DeclareAbbreviation\JBAA{J. Br. Astron. Assoc.}
\DeclareAbbreviation\JPhCS{J. of Physics Conference Series}
\DeclareAbbreviation\JPSJ{J. Phys. Soc. Japan}
\DeclareAbbreviation\JSARA{J. of the Southeastern Assoc. for Research in Astron.}
\DeclareAbbreviation\LowOB{Lowell Obs. Bull.}
\DeclareAbbreviation\MitAG{Mitteil. der Astronom. Gesell. Hamburg}
\DeclareAbbreviation\MitVS{Mitteil. Ver\"{a}nderl. Sterne}
\DeclareAbbreviation\MmSAI{Mem. Soc. Astron. Ital.}
\DeclareAbbreviation\memsai{Mem. Soc. Astron. Ital.}
\DeclareAbbreviation\Msngr{Messenger}
\DeclareAbbreviation\NewA{New Astron.}
\DeclareAbbreviation\na{New Astron.}
\DeclareAbbreviation\NewAR{New Astron. Rev.}
\DeclareAbbreviation\nar{New Astron. Rev.}
\DeclareAbbreviation\NInfo{Nauchnye Informatsii}
\DeclareAbbreviation\NPhS{Nature Physical Science}
\DeclareAbbreviation\OAP{Odessa Astron. Publ.}
\DeclareAbbreviation\Obs{Observatory}
\DeclareAbbreviation\OEJV{Open Eur. J. on Variable Stars}
\DeclareAbbreviation\PASA{Publ. Astron. Soc. Australia}
\DeclareAbbreviation\PASAu{Publ. Astron. Soc. Australia}
\DeclareAbbreviation\PAZh{Pis'ma AZh}
\DeclareAbbreviation\PJAB{Proc. Japan Acad. Ser. B}
\DeclareAbbreviation\POBeo{Publ. de l'Observatoire Astronomique de Beograd}
\DeclareAbbreviation\PCCP{Phys. Chem. Chem. Phys.}
\DeclareAbbreviation\PhR{Phys. Rep.}
\DeclareAbbreviation\PVSS{Publ. Variable Stars Sect. R. Astron. Soc. New Zealand}
\DeclareAbbreviation\PZ{Perem. Zvezdy}
\DeclareAbbreviation\PZP{Perem. Zvezdy, Prilozh.}
\DeclareAbbreviation\QJRAS{QJRAS}
\DeclareAbbreviation\RA{Ricerche Astronomiche}
\DeclareAbbreviation\RMxAA{Rev. Mexicana Astron. Astrof.}
\DeclareAbbreviation\RvMA{Reviews of Modern Astron.}
\DeclareAbbreviation\SASS{Society for Astronom. Sciences Ann. Symp.}
\DeclareAbbreviation\Sci{Science}
\DeclareAbbreviation\SPIE{SPIE Proc.}
\DeclareAbbreviation\SvA{Soviet Astronomy}
\DeclareAbbreviation\SvAL{Soviet Astronomy Letters}
\DeclareAbbreviation\VeSon{Ver\"{o}ff. Sternw. Sonneberg}
\DeclareAbbreviation\VSOLJBul{VSOLJ Variable Star Bull.}
\DeclareAbbreviation\yCat{VizieR Online Data Catalog}
\DeclareAbbreviation\ZA{Z. Astrophys.}

\def\ASPConf#1#2{ASP Conf. Ser. #1, #2}

\def\PublisherCambridge{Cambridge: Cambridge University Press}
\def\PublisherKluwer{Dordrecht: Kluwer Academic Publishers}
\def\PublisherASP{San Francisco: ASP}

\def\PublisherWorldScientific{Singapore: World Scientific Publishing}

\begin{document}
\SetRunningHead{T. Kato and Y. Osaki}{Three Dwarf Novae in the Kepler Field}

\Received{201X/XX/XX}
\Accepted{201X/XX/XX}

\title{Analysis of Three SU UMa-Type Dwarf Novae in the Kepler Field}

\author{Taichi \textsc{Kato}}
\affil{Department of Astronomy, Kyoto University,
       Sakyo-ku, Kyoto 606-8502}
\email{tkato@kusastro.kyoto-u.ac.jp}

\and

\author{Yoji \textsc{Osaki}}
\affil{Department of Astronomy, School of Science, University of Tokyo,
Hongo, Tokyo 113-0033}
\email{osaki@ruby.ocn.ne.jp}



\maketitle

\begin{abstract}
We studied the Kepler light curves of three SU UMa-type dwarf novae.
Both the background dwarf nova of KIC 4378554 and V516 Lyr showed
a combination of precursor-main superoutburst, during which superhumps
always developed on the fading branch of the precursor.  This finding
supports the thermal-tidal instability theory 
as the origin of the superoutburst. 
A superoutburst of V585 Lyr recorded by Kepler did not show a precursor 
outburst and the superhumps developed only after the maximum light, 
a first example in the Kepler data so far. 
Such a superoutburst is understood within the thermal-tidal instability model 
as a ``case B'' superoutburst discussed by \citet{osa03DNoutburst}.  
The observation of V585 Lyr made the first clear 
Kepler detection of the positive period derivative commonly seen 
in the ``stage B'' superhumps in dwarf novae with short orbital periods.
In all objects, there was no strong signature of a transition to 
the dominating stream impact-type component of superhumps.
This finding suggests that there is no strong
indication of an enhanced mass-transfer following the superoutburst.
In V585 Lyr, there were ``mini-rebrightenings'' with amplitudes of
0.2--0.4 mag and periods of 0.4--0.6~d between the superoutburst
and the rebrightening.  We have determined the orbital period
of V516 Lyr to be 0.083999(8)~d. In V516 Lyr, some of outbursts were 
double outbursts in a various degree. 
The preceding outburst in the double outburst was of the inside-out nature 
while the following one was of the outside-in nature. 
One of superoutbursts in V516 Lyr was 
preceded by a double precursor.  The preceding precursor failed to trigger 
a superoutburst and the following precursor triggered a superoutburst 
by developing positive superhumps. We have also developed
a method for reconstructing the superhump light curve and for
measuring the times of maxima from poorly sampled Kepler LC data.
\end{abstract}

\section{Introduction}

   Cataclysmic variables (CVs) are close binary systems consisting
of a white dwarf and a mass-transferring red dwarf star.
Among CVs, there are dwarf novae (DNe) which show semi-periodic
outbursts with amplitudes of 2--8 mag [for a review of CVs and DNe,
see e.g. \citet{war95book}].  Some of the DNe, called the SU UMa subtype, 
 show longer and brighter
outbursts called superoutburst, during which semi-periodic modulations
(superhumps) having periods one-to-several percent longer than
the orbital period ($P_{\rm orb}$).  The superhump period ($P_{\rm SH}$)
is believed to represent the synodic period between $P_{\rm orb}$
and the precession period of the non-axisymmetric (also referred
to as eccentric or flexing) accretion disk.  This non-axisymmetric
deformation of the disk is believed to be caused by the 3:1 resonance
tidal instability (\cite{whi88tidal}; \cite{hir90SHexcess};
\cite{lub91SHa}).  It is generally believed that the SU UMa-type
phenomenon is a combination of the thermal 
instability (\cite{osa74DNmodel}; \cite{mey81DNoutburst}) 
and tidal instability, and this is called the thermal-tidal
instability (TTI) model (\cite{osa89suuma}; see, \cite{osa96review} 
for a review).

   There have been, however, long-lasting debates whether enhanced
(or modulated) mass-transfer either plays a role in producing 
superoutbursts or superhumps [generally called enhanced mass-transfer
(EMT) model; e.g. \cite{sma91suumamodel}; \cite{sma04EMT}; 
\cite{sma08zcha}], or an enhanced mass-transfer caused by the outburst
may modify the behavior of the outburst (cf. \cite{las01DIDNXT};
\cite{pat02wzsge}), while it has been argued that an enhanced
mass-transfer is difficult to occur in an irradiated secondary star
in DNe (cf. \cite{osa04EMT}; \cite{via07irrsecondary};
\cite{via08irrmasstransfer}).

   By analysis of the Kepler data of V1504 Cyg, 
\citet{osa13v1504cygKepler} have demonstrated that the TTI model 
best explains the general behavior of SU UMa-type dwarf novae, 
particularly the disk radius variation during the supercycle 
(the cycle between the successive superoutbursts) by employing
the frequency of negative superhumps as a probe.  The data
presented in \citet{osa13v1504cygKepler} also indicated that
there is no pronounced enhanced mass-transfer before or around
the start of the superoutburst, disqualifying the EMT model
as a dominant mechanism to produce superoutbursts.  Furthermore,
\citet{osa13v1504cygKepler} demonstrated that the appearance 
of superhumps (manifestation of the tidal instability)
is closely related to the development of the superoutburst,
that is the development from the precursor outburst to the main
superoutburst, and they concluded that the superhump and superoutburst 
are so much entwined that one is almost difficult to find any 
interpretation other than the TTI model.  \citet{osa13v344lyrv1504cyg}
further studied Kepler observations of V1504 Cyg and V344 Lyr
by using a newly developed two-dimensional least absolute shrinkage 
and selection operator (Lasso) power spectra to strengthen
the conclusion.

   The situation for the enhanced mass-transfer caused by the outburst
is less clear.  There have been Kepler observations of V344 Lyr
(\cite{woo11v344lyr}; \cite{Pdot3}), which showed prominent 
secondary superhump maxima during the late stage of the superoutburst
which smoothly developed into so-called late superhumps, which
have $\sim$0.5 phase shift from main superhumps \citep{vog83lateSH}.
The strength of these secondary superhump maxima
might suggest an enhanced mass-transfer as their origin.
Since the strength of the secondary superhump was reported to be much
weaker in V1504 Cyg \citep{Pdot3}, it is necessary to study whether
the secondary superhump or late superhumps with $\sim$0.5 phase offset
are generally present in many SU UMa-type dwarf novae.  The ground-based
study seems to suggest that these late superhumps with $\sim$0.5 phase
offset are relatively rare except objects with high mass-transfer
rates ($\dot{M}$) \citep{Pdot3}.  This needs to be tested
by higher-quality observations.

   The presence of late superhumps with $\sim$0.5 phase shift in
low-$\dot{M}$ systems are also a point of debate since
a close examination of a large compilation of ground-based observations
suggests that most of previously reported late superhumps 
with $\sim$0.5 phase shift can be explained by a combination
of the discontinuous decrease of the superhump period and the gap
(traditional observations usually had gaps in the daytime
from one observing location) in the observation \citep{Pdot}.
This interpretation also requires confirmation by uninterrupted,
high-quality space observations.

   NASA's Kepler satellite (\cite{bor10Keplerfirst}; \cite{Kepler})
which was aimed to detect extrasolar planets, provide the best
opportunity to answer such questions.  The Kepler field contains
several CVs, and so far observations using the Kepler data 
have been reported for four SU UMa-type dwarf novae and one 
SS Cyg dwarf nova.  They are
V344 Lyr (\cite{sti10v344lyr}; \cite{can10v344lyr}; \cite{woo11v344lyr};
\cite{Pdot3}; \cite{can12v344lyr}; \cite{osa13v344lyrv1504cyg}),
V1504 Cyg (\cite{Pdot3}; \cite{can12v344lyr}; \cite{osa13v1504cygKepler};
\cite{osa13v344lyrv1504cyg}; \cite{coy12v1504cyg}),
V516 Lyr (\cite{gar11v516lyratel3507}),
the background dwarf nova of KIC 4378554 (\cite{bar12j1939}); and
an SS Cyg-type dwarf nova with shallow eclipses ---
V447 Lyr (\cite{ram12v447lyr}).
In addition to these DNe, other classes of CVs have been also
studied or discovered: MV Lyr (\cite{sca12shortvarmvlyr}; 
\cite{sca12mvlyr}), KIC 8751494 (\cite{kat13j1924}) 
(novalike variables), 
KIC 4547333 (\cite{fon11j1908}) (AM CVn-type object).
Kepler observations were either performed with high time-resolution
($\sim$1 min) mode, or short cadence (SC) runs or low time-resolution
($\sim$29.4 min) mode, or long cadence (LC) runs.  Since the number of
targets observable with SC is severely limited, most of Kepler
targets were recorded with the LC mode.  Some of these CVs were
also recorded with the SC mode.

   Although these objects are expected to play an important role
in studying superhumps in detail, there are several obstacles
which were not met in V1504 Cyg and V344 Lyr: faintness of
the objects (V516 Lyr, the background dwarf nova of KIC 4378554),
the presence of a close contaminating star (the background dwarf
nova of KIC 4378554), the lack of suitable SC runs (many objects).
The lack of suitable SC runs makes the analysis particularly 
difficult because the time-resolution of LC runs are typically
a third or a fourth of the superhump period of a typical SU UMa-type
dwarf nova.  There is a need to develop methods to tackle
these difficulties.  Here, we present the results of three SU UMa-type
dwarf novae using new techniques.

\section{Background Dwarf Nova of KIC 4378554}

\subsection{Introduction and Data Analysis}

   This SU UMa-type dwarf nova was discovered by \citet{bar12j1939}
as a background object of the Kepler target star of KIC 4378554.
We analyzed the public LC Kepler data to better characterize this object.

   Since the light of this object is mostly outside the aperture
of the original target star KIC 4378554 ({\tt SAP\_FLUX}), 
we applied a custom aperture
centered on this object just as in \citet{bar12j1939}.
Since we are mainly interested in the short-term variation
(i.e. superhumps) rather than the outburst amplitudes,
we did not adopt a sophisticated method described in \citet{bar12j1939}.
We added the observed count rates in the custom aperture
and subtracted for the ``sky'' values estimated from the surrounding
pixels just as in ordinary aperture photometry
(see appendix \ref{sec:handlepixel} for details).
Since the background is highly contaminated by the light of
KIC 4378554 and the systematic background level is slowly variable
\citep{bar12j1939}, we subtracted linear fits obtained from
the data outside the outbursts.  This procedure produces
the zero count for the averaged quiescent brightness of this object.
This is considered to be a good first-order approximation because this
object was undetectable in Kepler full-frame image in quiescence
\citep{bar12j1939}.  The numbers of pixels for the custom aperture
were three for Q3 and Q4, two for Q5 and Q7.  For quarters in which
the object was not detected in outburst, we used the apertures
used in the nearest quarters (three for Q1 and Q2, two for Q8).

   Since the superhump period is relatively long ($\sim$4 LC
exposures), we simply applied the fitting method described
in \citet{Pdot} to determine the times of superhump maxima
and the amplitudes as follows.
We extracted times of superhump maxima by numerically fitting
a template superhump light curve around the times of observed
maxima.  We used a template averaged (spline interpolated)
superhump light curve in GW Lib,
which shows a typical assymmetric profile of superhumps in
SU UMa-type dwarf novae.  We used two superhump cycles (typically
containing 6--8 data points) around maxima.
This method worked well, and it has become
evident that we do not have to pay special attention for
the poor time-resolution of LC measurements in characterizing
superhumps in long-$P_{\rm SH}$ systems.

\begin{figure*}
  \begin{center}
    \FigureFile(160mm,230mm){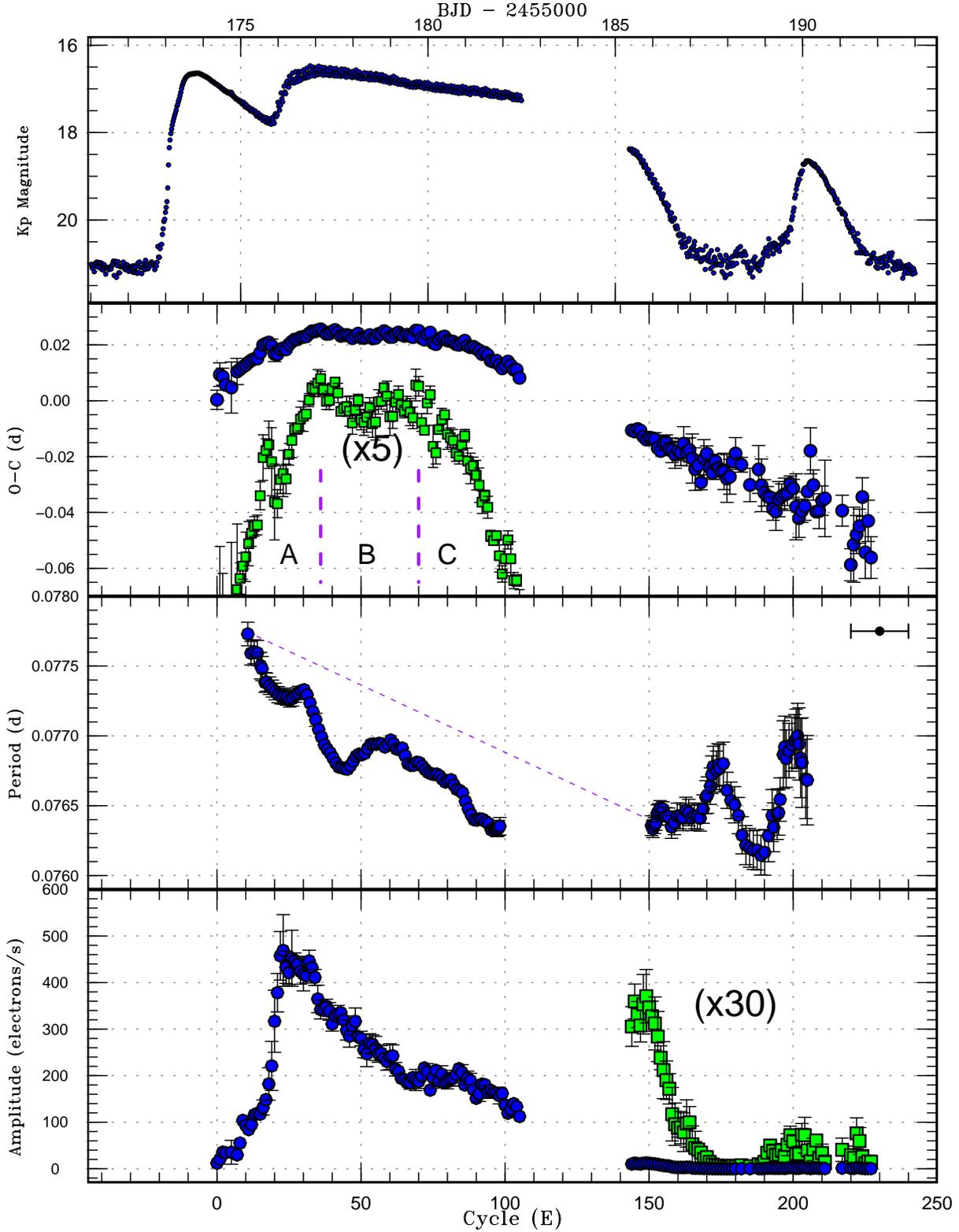}
  \end{center}
  \caption{Analysis of the background dwarf nova of KIC 4378554.
  From top to bottom:
  (1): Light curve.  A constant was added to the flux so that
  the outburst amplitude matches the typical amplitudes ($\sim$ 5 mag)
  of superoutbursts in SU UMa-type dwarf novae with similar $P_{\rm SH}$.
  This treatment is artificial and global trend of
  variation in the faintest part is not reliable.
  (2): $O-C$ diagram of the superhumps (filled circles).
  Filled squares represent $O-C$ values multiplied by five (and shifted
  arbitrarily by a constant) to better visualize the subtle variation
  during the superoutburst.
  A ephemeris of BJD $2455174.368+0.07690 E$ was used to draw this figure.
  (3): Period determined from the $O-C$ diagram.
  The period was determined by a linear regression to 
  the adjacent 20 times of maxima ($\sim$1.5~d).
  The window width is indicated by a horizontal bar at
  the upper right corner.
  The error refers to 1$\sigma$ error in this regression.
  Note that this process introduce artificial smoothing of
  the period variation and the error is only a random error
  which is likely smaller than the actual error.
  The dashed line is an imaginary global trend of the period 
  decrease in the absence of the pressure effect
  assuming that the period changes linearly
  (will be discussed in subsection \ref{sec:globaldecrease}).
  (4): Amplitudes of the superhumps.
  }
  \label{fig:nik1humpall}
\end{figure*}

   The result for the superoutburst and its associated normal
outbursts in Q3--Q4 is shown in figure \ref{fig:nik1humpall}.
In order to better produce the brightness variation during
outbursts, we added a constant to the flux (which was originally
adjusted to be zero outside the outbursts) so that
the quiescent magnitude matches the typical amplitudes ($\sim$5 mag)
of superoutbursts in SU UMa-type dwarf novae with similar $P_{\rm SH}$.
Although this treatment is artificial and the global trend of
variation in the faintest part is not reliable, 
this correction is expected to produce
a better representation of the light curve when the system
was in low brightness, especially the post-superoutburst stage.
We could not detect the orbital period in the quiescent data.
In calculating the superhump periods (figure \ref{fig:nik1humpall},
third panel), we made a linear regression to 20 adjacent times of
maxima measured by template fitting.
We should note that variation of the period is smoothed
by this time scale ($\sim$1.5~d).
The 1$\sigma$ error is a formal error of linear regression.

\subsection{System Characteristics}

   Although \citet{bar12j1939} quoted, with respect to this object,
VW Hyi, which is a prototype SU UMa-type dwarf nova with
frequent outbursts, we suggest that this object belongs to
a group of objects with much low $\dot{M}$
because typical intervals of normal outbursts (OB3 and OB4,
OB4 and OB5 in \cite{bar12j1939}) are an order of 200~d,
and more than 200~d quiescent interval was recorded before
OB1.  These recurrence times are much longer than the typical
recurrence times of 28~d in VW Hyi (e.g. \cite{vanderwoe87vwhyi}).
The long recurrence times recorded in this object is
more characteristic to SU UMa-type dwarf novae with
lower $\dot{M}$.  If we assume the cycle length
of normal outbursts ($T_n$) is proportional to $\dot{M}^{-2}$
(cf. \cite{ich94cycle}; \cite{osa96review}), $\dot{M}$ in V516 Lyr is
expected to be $\sim$2.7 times smaller than in VW Hyi
and with this $\dot{M}$ the superoutburst recurrence time 
(i.e., the supercycle length) is expected to be about 500--700~d.

   The sequence of outbursts shown in figure \ref{fig:nik1humpall}
is very characteristic to those of such low-$\dot{M}$ dwarf novae:
precursor outburst -- main superoutburst -- post-superoutburst
rebrightening.  Similar examples are found in many objects:
the 2003 superoutburst of V699 Oph \citep{Pdot}, 
the 2007 and 2009 superoutbursts of QZ Vir \citep{ohs11qzvir}.
Even when precursor outbursts were not present (this may have been
partly due to the lack of observations before the superoutbursts),
a post-superoutburst rebrightening is frequently observed
in many SU UMa-type dwarf novae (see examples in \cite{Pdot}),
and the precursor outburst is also a common feature in many
SU UMa-type dwarf novae (see e.g. \cite{kat97tleo}; \cite{uem05tvcrv}; 
\cite{she11j0423}; \cite{ima09j0532}; \cite{woo11v344lyr}).
We therefore identified
the sequence of outbursts (OB1--OB3 in \cite{bar12j1939})
as a complex of precursor outburst (OB1) -- main superoutburst (OB2)
-- post-superoutburst rebrightening (OB3).  
The short interval between the end of OB2 and OB3, 
as compared to those of other normal outbursts,  is then naturally 
understood.  Post-superoutburst rebrightenings, however, are more
frequently observed in shorter-$P_{\rm orb}$ systems
(cf. \cite{kat98super}).  There was a deep dip between
the precursor and the main superoutburst.  Such a deep dip
is difficult to explain in the pure thermal instability model
proposed by \citet{can10v344lyr}, because the precursor in their model 
just looks like a shoulder in the superoutburst light curve 
[cf. subsection 3.2 in \cite{osa13v1504cygKepler}].

   SU UMa-type dwarf novae with $P_{\rm SH}$ similar to this object
tend to show more frequent outbursts.  There are only few
moderately studied systems having comparable $P_{\rm SH}$ and 
outburst frequencies to this object.  We give an example of
EG Aqr [$P_{\rm SH}$=0.078828(6)~d; \cite{ima08egaqr}].
EG Aqr showed a superoutburst in 2006 with a small precursor
and the intervals of superoutburst were 800--1000~d
\citep{Pdot4}.  The supercycle length of EG Aqr is similar to 
what is expected ($\sim$500--700~d) for this object from the cycle
length of normal outbursts.
QY Per [$P_{\rm SH}$=0.07861(2)~d; \cite{Pdot}]
may be another similar object, although the frequency of outbursts
is lower in QY Per.
The representative objects having $P_{\rm SH}$
similar to this object but showing less frequent outbursts,
RZ Leo (\cite{ish01rzleo}) and V1251 Cyg (\cite{kat95v1251cyg};
\cite{Pdot2}) do not resemble the present object in that they
show intervals of several days (with double-wave modulations)
before superhumps appear and in that they do not show precursor
outbursts.

\subsection{Development of Superhumps}

   The second and bottom panels of figure \ref{fig:nik1humpall}
show the development of superhumps in this system.
As can be seen more clearly in the enlarged $O-C$ diagram
in the second panel, there are clear three distinct stages
as discussed in \citet{Pdot}: a stage with long, relatively
constant superhump period ($E \lesssim 30$; stage A),
a stage with a shorter period with a zero to slightly positive
period derivative ($P_{\rm dot} = \dot{P}/P$) ($30 \lesssim
E \lesssim 70$; stage B) and stage with a shorter superhump
period ($E \gtsim 70$; stage C).  As shown in figure 4
of \citet{Pdot}, objects with longer $P_{\rm SH}$ tend to
have shorter duration of stage B.  This tendency well matches
the behavior in this object.

   The transition from stage A to B,
however, took place later than when the amplitude of superhumps
reached a maximum.  Such a finding was not recorded in the
ground-based study, and it may be one of the advantages
in analyzing the Kepler data.  Using the epochs of superhump
maxima for $0 \le E \le 17$, before the amplitude of the
superhumps suddenly grew, we obtained a mean superhump period of
0.07781(10)~d.  We regarded this period to be the representative
superhump period when the superhumps were growing (here we call
it stage A1).  The latter part of stage A (we here call it
stage A2) has a mean period of 0.07727(5)~d, slightly shorter
than the period during stage A1.

\begin{figure}
  \begin{center}
    \FigureFile(88mm,110mm){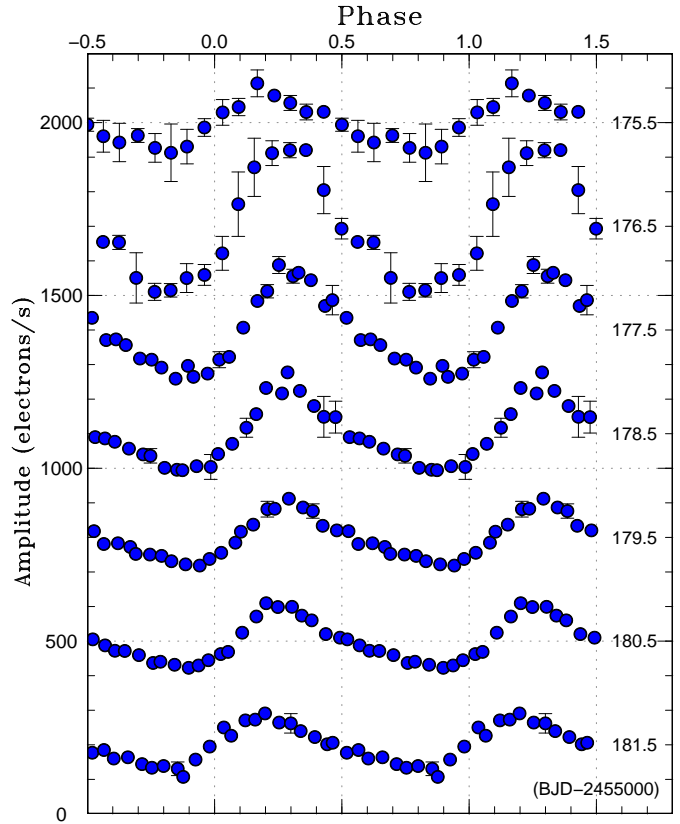}
  \end{center}
  \caption{Mean profiles of superhumps on individual 1-d segments
  during the early stage
  of the superoutburst of the background dwarf nova of KIC 4378554.
  The error bar represents a 1$\sigma$ error for the average
  of each phase bin.
  The phases were calculated by
  an ephemeris of BJD $2455174.368+0.07690 E$.
  The phases correspond to figure \ref{fig:nik1humpall}.
  }
  \label{fig:nik1prof1}
\end{figure}

   Using the segment of $36 \le E \le 70$, when the $O-C$ diagram
showed a positive $P_{\rm dot}$, we determined the mean period
of stage B superhumps as 0.07690(2)~d and $P_{\rm dot}$ as
$+13(4) \times 10^{-5}$.  Using the segment $70 \le E \le 105$,
we determined the period of stage C superhumps to be
0.07650(2)~d.  The period of stage C superhumps is 0.5\% shorter
than the mean period of stage B superhumps, in good agreement
with the general behavior described in \citet{Pdot}.
During stage B, the amplitude of superhumps monotonously
decreased while it grew again slightly after the transition
to stage C.  This behavior is also similar to other systems
\citep{Pdot3}.  The mean profiles of superhumps on individual 1-d
segments are shown in figure \ref{fig:nik1prof1}.

\begin{figure}
  \begin{center}
    \FigureFile(88mm,110mm){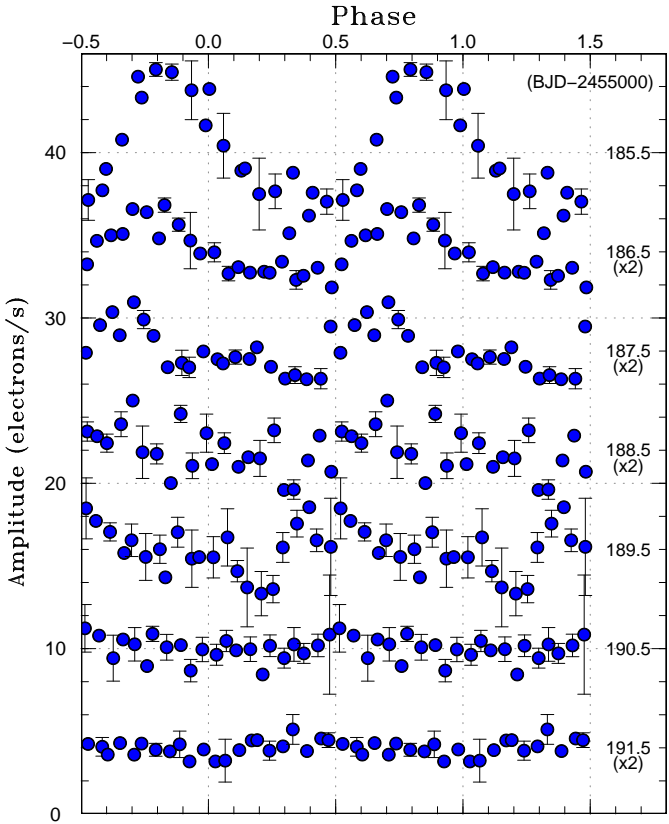}
  \end{center}
  \caption{Mean profiles of superhumps on individual 1-d segments
  during the late stage
  of the superoutburst, post-superoutburst and the rebrightening
  in the background dwarf nova of KIC 4378554.
  The error bar represents a 1$\sigma$ error for the average
  of each phase bin.
  The phases were calculated by
  an ephemeris of BJD $2455174.368+0.07690 E$.
  The phases correspond to figure \ref{fig:nik1humpall}.
  The negative phases for the maxima are simply a result
  of the adoption of the constant epoch and period.
  The smooth variation of the phase of maxima indicates
  the absence of a phase shift.
  }
  \label{fig:nik1prof2}
\end{figure}

   Although there was an unavoidable gap (a gap between Kepler
quarters) after $E=105$, the later ($E \ge 144$)
development in the $O-C$ diagram when Kepler restarted observation appears to
be on a smooth extension of stage C superhumps.
This suggests that the period and phase of superhumps did not 
greatly change during the gap of the observations.
By using all $E \ge 70$ times of maxima, we obtained a period
of 0.07643(1)~d.  The object immediately entered the rapid
fading stage of the superoutburst as the observation resumed.
The $O-C$ curve did not show a strong variation despite the
rapid brightness decline.  This feature was also commonly seen
in many SU UMa-type dwarf novae (e.g. \cite{Pdot}).
The amplitudes of superhumps also decreased as the system faded
(see also figure \ref{fig:nik1prof2}),
and this makes clear difference to persisting superhumps
in V344 Lyr (\cite{woo11v344lyr}; \cite{osa13v344lyrv1504cyg}),
which are supposed to arise from the accretion stream-disk
interaction, corresponding to the traditional picture of late
superhumps \citep{vog83lateSH}.  
We should note that the enhancement of the secondary hump in 
figure 9 of \citet{bar12j1939} corresponds to the late
post-superoutburst stage (corresponding to day 189.5, where 
day $x$ represents BJD 2455000$+x$) in figure \ref{fig:nik1prof2}).
There is some hint of the enhancement
of the superhump amplitudes (measured in electrons s$^{-1}$)
during the post-superoutburst rebrightening.

   As already discussed in \citet{Pdot} and
in subsequent series of papers, the late stage (post-superoutburst)
superhumps in low-$\dot{M}$ SU UMa-type dwarf novae bear
more characteristics of continuation of stage C superhumps
than stream-disk interaction-type traditional late superhumps.
This finding in the high sensitivity Kepler data adds another
support that the mass-transfer rate is not greatly enhanced
during superoutbursts [see also a discussion in \citet{osa03DNoutburst}].

\subsection{Oscillations before the Post-Superoutburst Rebrightening}
\label{sec:j1939osci}

   In the light curve, there appears to be some signature of
oscillations between the superoutburst and the post-superoutburst
rebrightening.  There are small bump-like structures around
BJD 2455188.3 and 2455189.3 (the latter becomes more evident
after subtracting the rising trend of the rebrightening).
A trace of these features can be also seen in figure 5
of \citet{bar12j1939} and we consider that the feature is real.
Since the similar feature can be better seen in V585 Lyr,
we describe the characteristics in the section of V585 Lyr.

\section{V585 Lyrae}

\subsection{Introduction and Kepler Data}

   V585 Lyr is a dwarf nova discovered by \citet{kry01v585lyrv587lyr}.
\citet{kry01v585lyrv587lyr} reported a long outburst and two short
outbursts, and suggested that the object is an SU UMa-type
dwarf nova.  It is interesting that the long outburst showed
a temporary fading near the start of the outburst.  It was most
likely accompanied by a precursor.  \citet{Pdot} studied the 2003
superoutburst and \citet{Pdot4} reported on the less observed 
2012 superoutburst.  \citet{how13KeplerCVs} reported an unsuccessful
attempt to obtain a spectrum.

   The Kepler public data for V585 Lyr available in this writing are those of 
Q2--Q10 for LC and Q9 and Q14 for SC.    
V585 Lyr is sufficiently isolated from nearby stars,
we used {\tt SAP\_FLUX} for SC and LC data.  Within these data there was only 
one outburst event recorded by Kepler, 
a superoutburst and its post-superoutburst rebrightening,  
starting at the end of 2010 January and ending
at the end of February.  Only LC data are available during this 
superoutburst, and there was more than a 4-d gap immediately after
the peak of the superoutburst.  Due to these restrictions,
the early development of the superhumps was not sufficiently
recorded.  We can say, however, that there was no precursor outburst
and that low-amplitude modulations already appeared $\sim$6 hr after
the peak brightness.  These modulations likely had long periods
(several hours) around their appearance, but there was an indication
of emerging signals (BJD 2455229.4) having a period close to 
the superhump period, which we will discuss later.  Unfortunately, 
the Kepler observation was  stopped soon after these signal appeared.
We therefore mainly focus on the analysis after the resumption
of the observation at BJD 2455233.8.

\begin{figure}
  \begin{center}
    \FigureFile(88mm,110mm){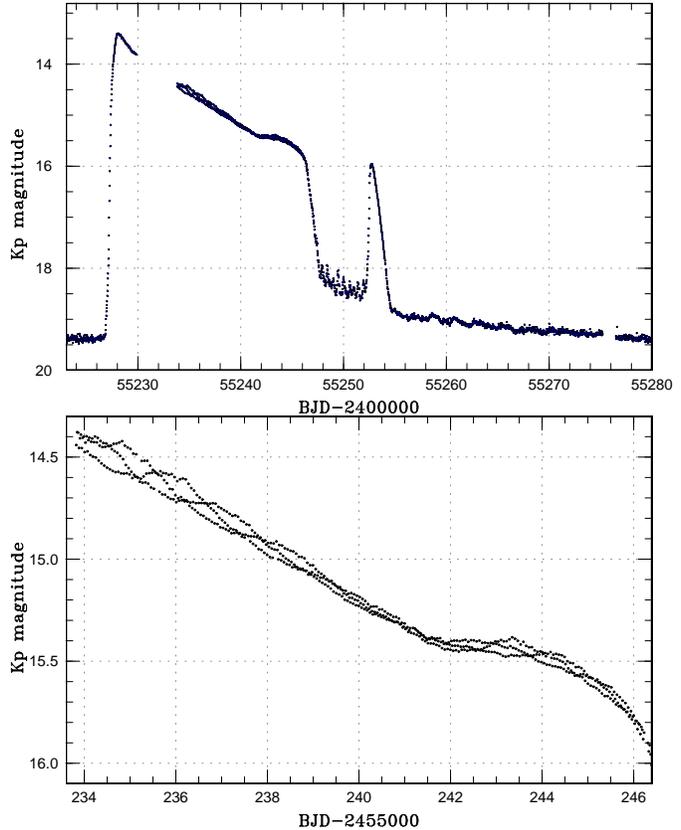}
  \end{center}
  \caption{The 2010 superoutburst of V585 Lyr in Kepler LC data.
  (Upper:) Entire outburst.
  (Lower:) Enlargement of the outburst to illustrates the low
  sampling rate of the LC data.  There appear three independent
  curves due to the beat between the three times the LC sampling
  and the superhump period.}
  \label{fig:v585out}
\end{figure}

\begin{figure}
  \begin{center}
    \FigureFile(88mm,60mm){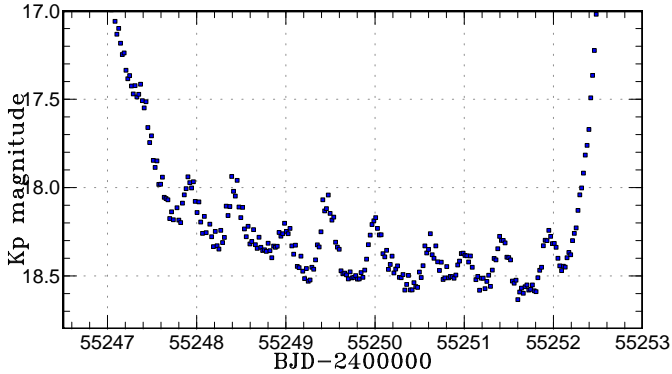}
  \end{center}
  \caption{The post-superoutburst stage of V585 Lyr
  with semi-periodic variations (``mini-rebrightenings'').}
  \label{fig:v585post}
\end{figure}

   The overall light curve of this superoutburst is shown in
figure \ref{fig:v585out}.  The superoutburst started around BJD 2455227 
and ended around BJD 2455248, thus its duration was about 21~d. 
It was followed by
the post-superoutburst stage with semi-periodic variations
(``mini-rebrightenings'': BJD 2455247.5--2455252; figure
\ref{fig:v585post}) and a distinct post-outburst rebrightening 
peaking at around BJD 2455252.7.
The object then faded slowly with semi-periodic
fluctuations with periods of $\sim$2~d and amplitudes of
0.1--0.2 mag.  These fluctuations almost ceased after the
following 20~d (BJD 2455276).

   The SC runs (Q9 and Q14) did not record any outbursts.
Although the object was closely observed in quiescence in 
two SC runs, we could not detect any significant periodic
signal for each SC run.  The analysis of LC data could neither
detect any signature of orbital variation.

\subsection{Variation of Superhump Profile from LC data}
\label{sec:v585prof}

   In order to obtain an $O-C$ diagram or the period variation
(subsection \ref{sec:v585oc}), we need the mean profile of
the superhump beforehand.  We therefore first obtain
the mean superhump profile and then discuss the $O-C$ variation.

   As discussed below, we can see periodic light variation of 
superhump origin during the superoutburst of V585 Lyr with 
a period of about 0.060~d (or 87~min or 16.5~c/d), which is very close to 
three times of the LC sampling interval of 29.42~min.
It is particularly difficult to obtain times
of maxima or the superhump profile by conventional methods due to 
this low sampling rate.  We therefore reconstructed the template
light curves for each 0.5~d bin using the Markov-chain Monte Carlo
(MCMC) modeling of the Kepler data (see appendix \ref{sec:recon}).

\begin{figure}
  \begin{center}
    \FigureFile(80mm,160mm){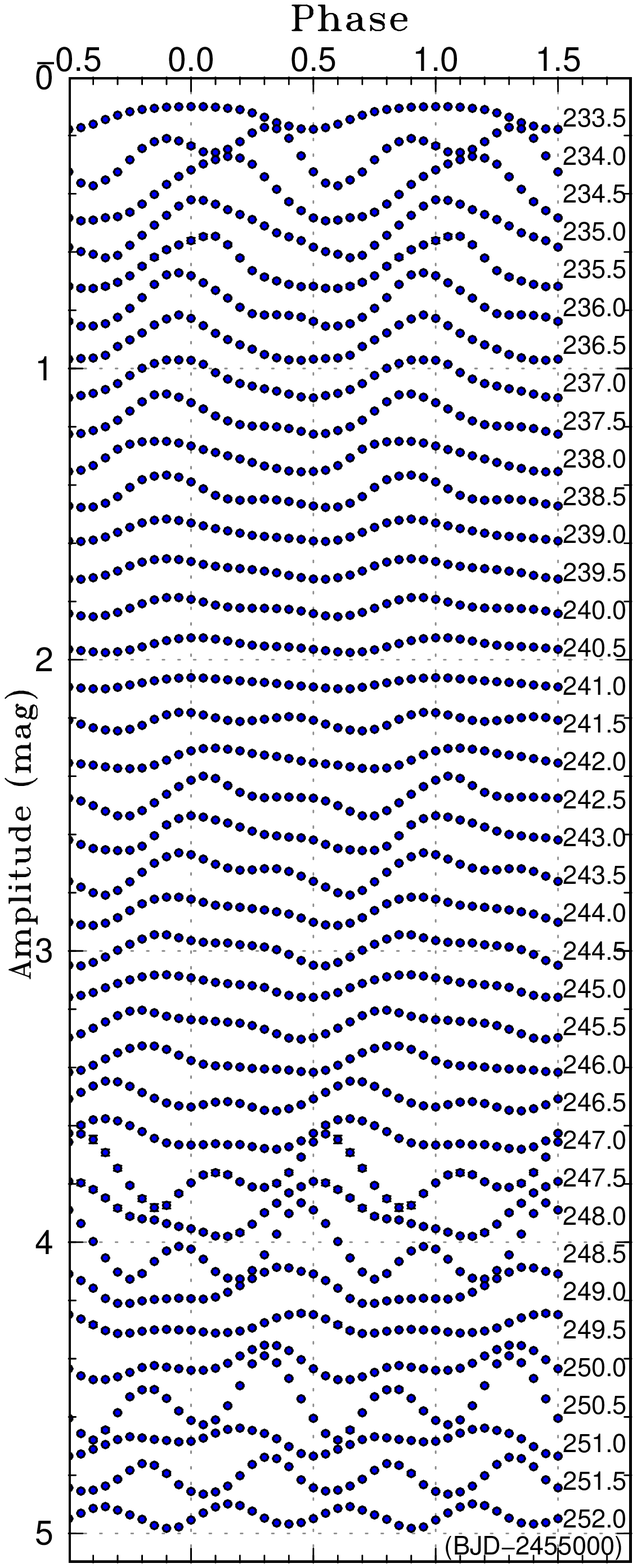}
  \end{center}
  \caption{Profile variation of superhumps in V585 Lyr.
  The profile for each 0.5-d segment of the LC data was estimated
  by modeling the observation (see text).  The errors (smaller than
  figure symbols) are not true errors but numerical 1 $\sigma$
  range of the probability distribution function.
  The phases were defined by an element of 
  BJD(max)$=2455233.7951+0.060437 E$.
  The profile was double-peaked at the end of stage A (day 234.0).
  The period increased during stage B (days 234.5--242.5), and secondary
  humps only appeared near the end of this stage (days 241.5--242.5).
  Stage C with a shorter superhump period was recorded during
  the later stage of the superoutburst and the subsequent fading stage.
  The day 246.0 corresponds to the start of the rapid fading.
  Despite this fading, no alternation between the main and secondary
  humps was observed as in V344 Lyr \citep{woo11v344lyr}.
  This figure is shown in amplitude scale for easier visibility.
  }
  \label{fig:v585prof}
\end{figure}

   The result is shown in figure \ref{fig:v585prof}.
The profile looked like double-peaked at the end of stage A (day 234.0).
This feature may have
been spurious since the profile was not double-peaked 
in the incomplete (shorter than 0.5~d) preceding bin (day 233.5).
The period increased during stage B (days 234.5--242.5), and secondary
humps only appeared near the end of this stage (days 241.5--242.5).
The amplitudes of the superhumps grew again.
Stage C with a shorter superhump period was recorded during
the later stage of the superoutburst and the subsequent fading stage.
The day 246.0 corresponds to the start of the rapid fading.
Despite this fading, no alternation between the main and secondary
humps was observed as in V344 Lyr \citep{woo11v344lyr}.
After day 247.5, the object entered the post-superoutburst stage
(before the rebrightening).  During this stage, mini-brightenings
with periods of 0.5--0.7~d and amplitudes of $\sim$0.3 mag
were recorded.  This phenomenon will be discussed later.
The superhumps became undetectable around
the peak of the rebrightening, and the waveform became 
difficult to trace after this rebrightening (not shown in the figure).

   This result well reproduces the ground-based observations
of short-$P_{\rm orb}$ systems such as SW UMa (\cite{soe09swuma};
\cite{Pdot}) and V585 Lyr itself \citep{Pdot}, although 
the ground-based observation only recorded a part of a superoutburst
of V585 Lyr with limited accuracy.

\subsection{$O-C$ Analysis of LC data}\label{sec:v585oc}

\begin{figure*}
  \begin{center}
    \FigureFile(160mm,230mm){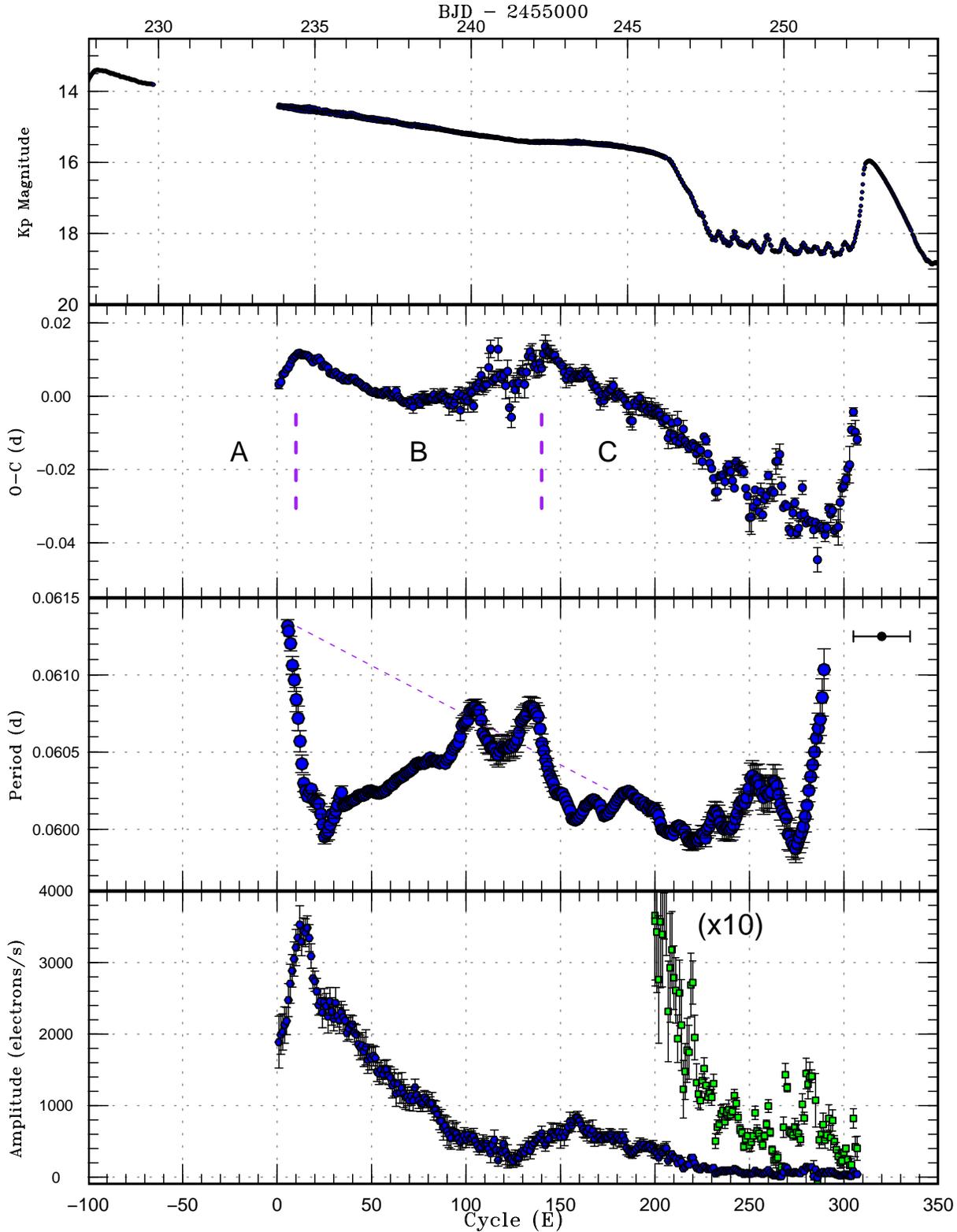}
  \end{center}
  \caption{$O-C$ diagram of V585 Lyr.
  From top to bottom:
  (1): Kepler LC light curve.
  (2): $O-C$ diagram.
  The figure was drawn against a period of 0.06043~d.
  (3): Period determined from the $O-C$ diagram.
  The period was determined by a linear regression to 
  the adjacent 30 times of maxima ($\sim$1.8~d).
  The window width is indicated by a horizontal bar at
  the upper right corner.
  During stage A, 10 ($\sim$0.6~d) adjacent maxima were used.
  The error refers to 1$\sigma$ error in this regression.
  Note that this process introduce artificial smoothing of
  the period variation and the error is only a random error
  which is likely smaller than the actual error.
  Wiggles (especially late stage B and stage C)
  were probably caused by the ``beat'' because the sampling rate 
  was almost exactly one third of the superhump period, 
  and they should not be considered as real.
  The dashed line is an imaginary global trend of the period 
  decrease in the absence of the pressure effect
  assuming that the period changes linearly
  (will be discussed in subsection \ref{sec:globaldecrease}).
  (4): Amplitudes in electrons s$^{-1}$.  Since modeling the
  Kepler data was performed by variable superhump template, the
  normalization of these amplitudes are not the same in different
  epochs.  The figure is presented to show the general trend.}
  \label{fig:v585humpall}
\end{figure*}

   Using the method described in appendix \ref{sec:maxdet},
we determined the times of superhump maxima
(figure \ref{fig:v585humpall}).  We here define $E=0$ for
BJD 2455233.8536.  The result very well demonstrates
the familiar stage A--B--C variations in the $O-C$ diagrams
in short-$P_{\rm orb}$ systems [cf. \citet{Pdot}, particularly
figure 3 for SW UMa ($P_{\rm orb}$=0.056815~d)]:
there was stage A with a long superhump period and with evolving
superhumps ($E \lesssim 10$), stage B with the increasing superhump
period ($10 \lesssim E \lesssim 140$) and stage C with a shorter,
and almost constant superhump period ($E \gtsim 140$).
The mean periods for these stages were 0.06128(4)~d (stage A),
0.06041(1)~d (stage B) and 0.06013(1)~d (stage C, limited to
$140 \le E \le 220$).  The $P_{\rm dot}$ for stage B was
$+9.6(5) \times 10^{-5}$, which is a very characteristic value
for this $P_{\rm SH}$ (\cite{Pdot}; \cite{Pdot2}; \cite{Pdot3};
\cite{Pdot4}).  These values very well agree the ground-based
value for the 2003 superoutburst: 0.06113(8)~d (stage A),
0.06036(2)~d (stage B), and $P_{\rm dot}$ = $+10.7(12) \times 10^{-5}$
for stage B.  The period of stage C superhump was not well determined
in 2003 due to the short observational coverage \citep{Pdot}.
The value of 0.06035(4)~d for stage B superhumps was obtained
from a more sparse data set in the 2012 superoutburst \citep{Pdot4}.
Although there was a possible rapid increase in the superhump
period around the rising phase the rebrightening ($E \ge 290$),
the reality of this phenomenon was difficult to confirm because
of the signal close to the detection limit.

\begin{figure}
  \begin{center}
    \FigureFile(88mm,100mm){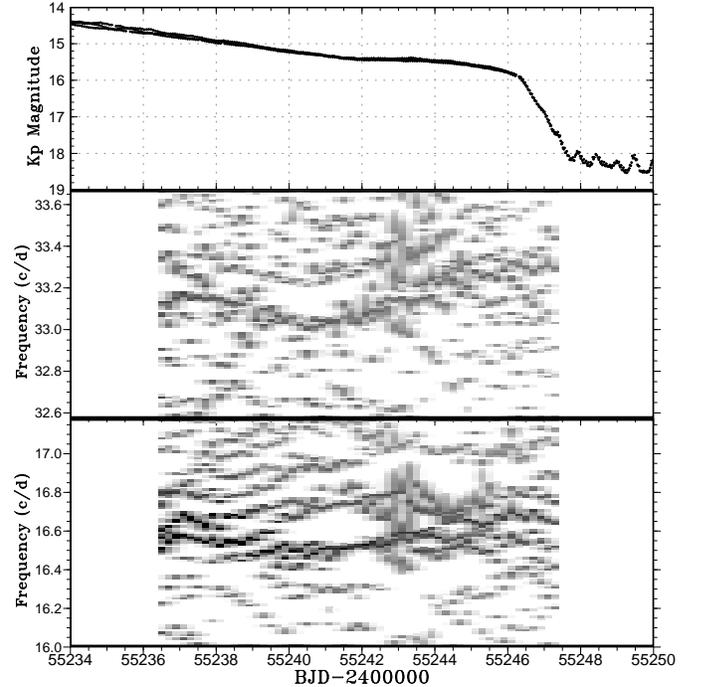}
  \end{center}
  \caption{Lasso analysis of superhumps of V585 Lyr in Kepler LC data.
  (Upper:) Light curve.
  (Middle:) First harmonics of the superhump signal.
  (Lower:) Fundamental of the superhump signal.
  The frequency of the strongest (superhump) signal initially decreased
  (up to BJD 2455241) and increased after then both in
  the fundamental and first harmonics.  This trend confirmed
  the result from the $O-C$ analysis.
  Note that the frequency of the first harmonics is higher than
  the Nyquist frequency, and the Lasso analysis can detect such
  a signal using appropriate windows.
  $\log \lambda=-5.3$ was used.  The width of 
  the sliding window and the time step used are 5~d and 0.2~d,
  respectively.
  }
  \label{fig:v585lyrspec2dlasso}
\end{figure}

   Independent confirmation of this pattern of period variation
was obtained by two-dimensional Lasso spectrum 
(figure \ref{fig:v585lyrspec2dlasso}).  This figure corresponds
to figure 8 in \citet{bar12j1939}, who used Fourier analysis.
The frequency of the strongest (superhump) signal initially decreased
(up to BJD 2455241; corresponding to stage B) and increased after
then (stage B--C transition) both in the fundamental and first harmonics.
This trend (i.e. initial increase and then decrease in the period) 
confirmed the result from the $O-C$ analysis.
Note that the frequency of the first harmonics is higher than
the Nyquist frequency.  The Lasso analysis can handle such a case
using appropriate windows and suppressing the reflected signal
at the Nyquist frequency, which was seen in \citet{bar12j1939}.

   The $O-C$ analysis of the Kepler LC data thus perfectly
confirmed the trend in short-$P_{\rm orb}$ systems recorded
in ground-based observations.  This object is the first Kepler
CV showing the very distinct stages A--B--C, particularly the clear
presence of stage B with an unambiguously positive $P_{\rm dot}$.
The Kepler observation, however, recorded the details of
stage C in unprecedented quality.  This observation unambiguously
demonstrated the lack of phase $\sim$0.5 jump (corresponding to
the ``traditional'' late superhumps).

   The variation of the superhump amplitudes was similar to
those in short-$P_{\rm orb}$ systems recorded in ground-based
observations: the peak of the superhump amplitude is close to
the stage A--B transition, the amplitude decreases during stage B
and reaches a minimum before stage B--C transition, and again
increases near the stage B--C transition \citep{Pdot3}.
In conjunction with the amplitude increase near the stage B--C 
transition, the brightness of the system also shows an upward 
(brighter) deviation from the linear fading.  This brightening
trend is commonly seen in short-$P_{\rm orb}$ systems
\citep{kat03hodel}.

\subsection{Mini-rebrightenings}

   Between the superoutburst and the distinct post-superoutburst
rebrightening (BJD 2455247.5--2455252; figure \ref{fig:v585post}),
there were ``mini-rebrightenings'' with amplitudes of 0.2--0.4 mag
and periods of 0.4--0.6~d.  As far as we know, such variations have
never been documented before.  The amplitudes appears to be too
large to be explained by a beat between some period and another
(e.g. between the superhump period and another period), and the
periods of 0.4--0.6~d appear to be too short to be explained
by a beat phenomenon.  We thus regard them as real brightness variation
with these time-scales.  There was a hint of the same phenomenon
on the declining branch from the superoutburst (BJD 2455247.4).
This phenomenon disappeared as the object experienced the distinct
rebrightening peaking at BJD 2455252.7, and never appeared again
after the fading from the rebrightening.

   Although this phenomenon appears to be physically related to 
the appearance of the post-superoutburst rebrightenings, and potentially
important to understand the origin of the rebrightening,
we still do not have a physical explanation for this phenomenon.
Not all SU UMa-type dwarf novae show post-superoutburst rebrightenings
\citep{Pdot}, and a systematic search for such variations might
lead to a solution.

   As we have already seen, the background dwarf nova of KIC 4378554
also possibly showed the same phenomenon (subsection \ref{sec:j1939osci}),
such a phenomenon may be more prevalent and will deserve
a further study.

\section{V516 Lyrae}

\subsection{Introduction}

   V516 Lyr was originally identified as a blue object (NGC 6791 B8)
in the region of an old open cluster NGC 6791 \citep{kal92ngc6791}.
\citet{kal95ngc6791} suspected that the object is likely a cataclysmic 
variable based on the color and variability.
The DN-type nature was confirmed by photometry and spectroscopy
by \citet{kal97v516lyrv523lyr}.
Ground-base observations indicated that this object showed $\sim$2 mag
outbursts outside the $V=21$ quiescence and was suggested to be
an SS Cyg-type dwarf nova \citep{moc03v516lyr}.
\citet{gar11v516lyratel3507} reported from Kepler SC data
that this object underwent a superoutburst on 2011 October 13
and determined the superhump period of 2.097(3) hr (in average) or
2.109(3) hr (initial two days) \citep{gar11v516lyratel3507}.
\citet{how13KeplerCVs} presented a spectrum in agreement with
the dwarf nova-type classification and also presented
Kepler light curves showing normal outbursts with the $\sim$18~d
recurrence time.  \citet{how13KeplerCVs}\footnote{
   In \citet{how13KeplerCVs}, V516 Lyr actually refers to V523 Lyr
   and V523 Lyr actually refers to V516 Lyr.  Their fiugre 4
   actually shows the lightcurve of V516 Lyr, not V523 Lyr.
} also reported
the possible detection of a periodicity of 0.087478~d by the LC 
data in the quiescent interval around BJD 2455468.

\subsection{Global Light Curve}\label{sec:v516lyrglobal}

\begin{figure*}
  \begin{center}
    \FigureFile(160mm,230mm){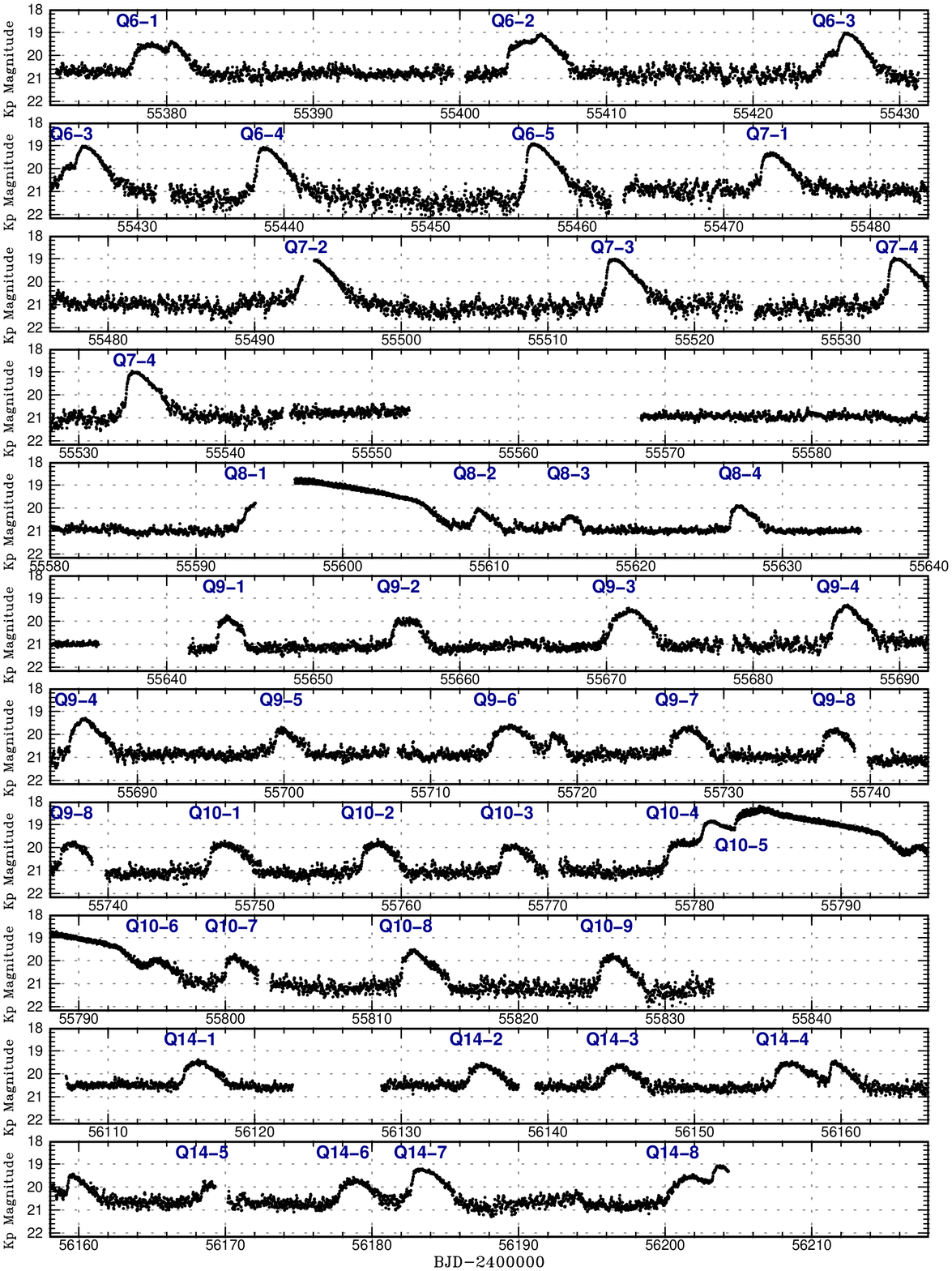}
  \end{center}
  \caption{The Kepler LC light curve of V516 Lyr.
  The numbers of the outbursts correspond to table \ref{tab:v516lyrout}.
  }
  \label{fig:v516lc}
\end{figure*}

   We first examine the LC Kepler light curves of V516 Lyr available 
for public at the present writing.  They are those of Q6--Q10 and Q14. 
Figure \ref{fig:v516lc} illustrates the Kepler light curve of 
the LC data of V516 Lyr for periods of Q6--10 and Q14.
Since there is no long-term stable zero-point in the Kepler
light curve, we artificially added constants to adjust the quiescent
level to be $\sim$21 mag, as recorded in \citet{moc03v516lyr}.
As seen from figure \ref{fig:v516lc}, V516 Lyr exhibited about 40 outbursts 
including two superoutbursts in a time scale of about one and a half years. 
We summarize the main characteristics of these outbursts 
in table \ref{tab:v516lyrout}.  The first 
column of the table is a sequential number of outbursts 
within a given quarter.  The second and third columns are the dates 
of the start and the end of an outburst (determined by eye), 
respectively, where dates are counted 
from BJD 2455000.  The fourth column is the outburst duration and the fifth 
column gives the outburst type. The sixth column gives the duration of 
quiescence preceding to a given outburst and the last column is just 
comments, if there are any. 

   As discussed by \citet{sma84DI}, two types of normal outbursts 
are recognized; Type A outburst or the ``outside-in'' outburst in which the 
heating transition of the thermal instability starts from the outer-part 
of the disk and the heating front propagates inward 
and Type B outburst or the ``inside-out'' outburst in which 
the heating transition does in the inner-part of the disk and the heating 
wave propagates outward. The light curve of the Type A outburst is 
characterized by a rapid rise to outburst maximum 
as compared to the slower decay from maximum 
while the type B outburst is characterized by more or less 
symmetrical rise and fall around maximum. The judgement of 
outburst types is made by eye and we must admit 
there are some delicate and ambiguous cases in this judgement.

   Two superoutbursts are seen in our data, which occurred around 
BJD 2455600 (SO No. 1) and 2455785 (SO No. 2). 
Thus the length of supercycle between these two superoutbursts is about 185~d. 
However, since the starting date of observations in Q6 was BJD 2455372 
and since no superoutburst was apparently missed before the SO No. 1, 
the length of the previous supercycle must be longer than 220~d at least. 
The next superoutburst is expected to have occurred most likely 
in the period of Q12 but so far we have no way to confirm or disprove. 

   As discussed in the next subsection, a possible orbital period 
of V516 Lyr was found to be 0.0840~d (2.016~hr or 11.9 c/d) 
which is very near to that of V344 Lyr. 
However the recurrence cycles of both the normal 
outburst and the superoutburst suggest that V516 Lyr has slightly lower 
mass-transfer rate than that of V344 Lyr. 

   One of the most interesting characteristics in the light curve of V516 Cyg 
was an occurrence of several double outbursts; their examples  are 
outbursts Q6-1, Q6-2, Q6-3, Q9-6, Q14-4, and Q14-8. The two outbursts of 
Q14-6 and Q14-7 are most likely a double outburst 
with a just one day quiescence interval. 
The degree of doubling varies from outburst to outburst, such as an outburst 
with a  shoulder (e.g., Q6-3) to an outburst with a deep dip almost 
touching quiescence (e.g., Q14-4). As seen in figure \ref{fig:v516lc}, 
the preceding outburst in the double 
outburst was always type B (``inside-out'') outburst 
while the following one was type A (``outside-in'') outburst and there 
is no exception to this rule so far. 
The double outbursts tended to occur in group, e.g., in Q6 and Q14. 

   This phenomenon is understood in the following way.  As mentioned above the 
preceding outburst is the inside-out outburst in which the heating transition 
first occurs in the inner part and the heating front propagates outward. But 
the heating front fails to reach the outer edge of the disk but is 
reflected in the middle of the disk (i.e., type Bb outburst 
in Smak's classification).  This creates a special situation 
in which the thermal instability is easily triggered in the outer part of 
the disk because a large amount of matter pushed outward by the heating front 
is left over in the outer-part by reflection (see, \cite{can93DIreview}). 
The second thermal instability in the outer part of the disk is then 
triggered (type A outburst) before the cooling wave of the preceding 
thermal instability completely reached the inner edge of the disk.  
The behavior of the double outburst can be well explained in the framework 
of the thermal disk instability.

\begin{table}
\caption{Properties of outbursts of V516 Lyr.}
\label{tab:v516lyrout}
\begin{center}
\begin{tabular}{ccccccc}
\hline
No. & start\commenta & end\commenta & $d_{\rm out}$\commentb & type\commentc & $d_{\rm qui}$\commentd & C\commente \\ 
\hline
Q6-1 & 377 & 381.6 & 4.6 & double & -- \\
Q6-2 & 402.6 & 407.5 & 4.9 & double & 21 \\
Q6-3 & 424 & 428.4 & 4.4 & shoulder & 16.5 \\
Q6-4 & 437.5 & 440.8 & 3.3 & A & 9.1 \\
Q6-5 & 456 & 460 & 4 & A & 15.2 \\
Q7-1 & 472 & 475 & 3 & A & 12 \\
Q7-2 & 492.5 & 496 & 3.5 & ? & 17.5 & 1 \\
Q7-3 & 513.4 & 516.8 & 3.4 & A & 17.4 \\
Q7-4 & 532.8 & 536 & 3.2 & A & 16 \\
Q8-1 & 592.4 & 606.8 & 14.4 & SO & 56? & 1\\
Q8-2 & 608 & 610.5 & 2.5 & RB & 1.2 \\
Q8-3 & 614.4 & 616 & 1.6 & mini & 3.9 \\
Q8-4 & 626.1 & 628.5 & 2.4 & A & 10.1 \\
Q9-1 & 642.9 & 645 & 2.1 & B & 14.4 & 1 \\
Q9-2 & 654.8 & 657.8 & 3 & B? & 9.8 & 2 \\
Q9-3 & 669.5 & 673.1 & 3.6 & B & 11.7 \\
Q9-4 & 684.8 & 688.2 & 3.4 & B & 11.7 \\
Q9-5 & 698.8 & 701.1 & 2.3 & A & 10.6 \\
Q9-6 & 713.4 & 719 & 5.6 & double & 12.3 \\
Q9-7 & 725.8 & 729 & 3.2 & B & 6.8 \\
Q9-8 & 736 & -- & -- & B & -- \\
Q10-1 & 746.3 & 749.9 & 3.6 & A? & -- \\
Q10-2 & 756.7 & 760 & 3.3 & B & 6.8 \\
Q10-3 & 766.2 & 769.2 & 3 & A & 6.2 \\
Q10-4 & 777.5 & 782.2 & 4.7 & double & PC \\
Q10-5 & 782.2 & 794 & 11.8 & SO & 0 \\
Q10-6 & 794 & 796.7 & 2.7 & RB & 0 \\
Q10-7 & 799.4 & 802 & 2.6 & A & 2.7 \\
Q10-8 & 811.3 & 814.8 & 3.5 & A & 9.3 \\
Q10-9 & 825 & 828 & 3 & A & 10.2 \\
Q14-1 & 1114.6 & 1117.9 & 3.3 & B & -- \\
Q14-2 & 1134.1 & 1137.4 & 3.3 & B & 16.2 \\
Q14-3 & 1143.4 & 1146.6 & 3.2 & B & 6 \\
Q14-4 & 1155 & 1161 & 6 & double & 8.4 \\
Q14-5 & 1167.4 & 1170 & 2.6 & mini & 6.4 \\
Q14-6 & 1177 & 1181.3 & 4.3 & B & 7 \\
Q14-7 & 1182.3 & 1185.8 & 3.5 & A & 1 \\
Q14-8 & 1199.5 & -- & -- & double? & 13.7 \\
\hline
  \multicolumn{6}{l}{\commenta BJD$-$2455000.} \\
  \multicolumn{6}{l}{\commentb Outburst duration (d).} \\
  \multicolumn{6}{l}{\parbox{230pt}{\commentc Abbreviations: precursor (PC),
  rebrightening (RB), superoutburst (SO).}} \\
  \multicolumn{6}{l}{\commentd Quiescence duration prior to the outburst (d).} \\
  \multicolumn{6}{l}{\commente Comment. 1: data gap 2: too noisy.}\\
\end{tabular}
\end{center}
\end{table}

\subsection{SC Data}\label{sec:v516lyrsc}

\begin{figure}
  \begin{center}
    \FigureFile(88mm,110mm){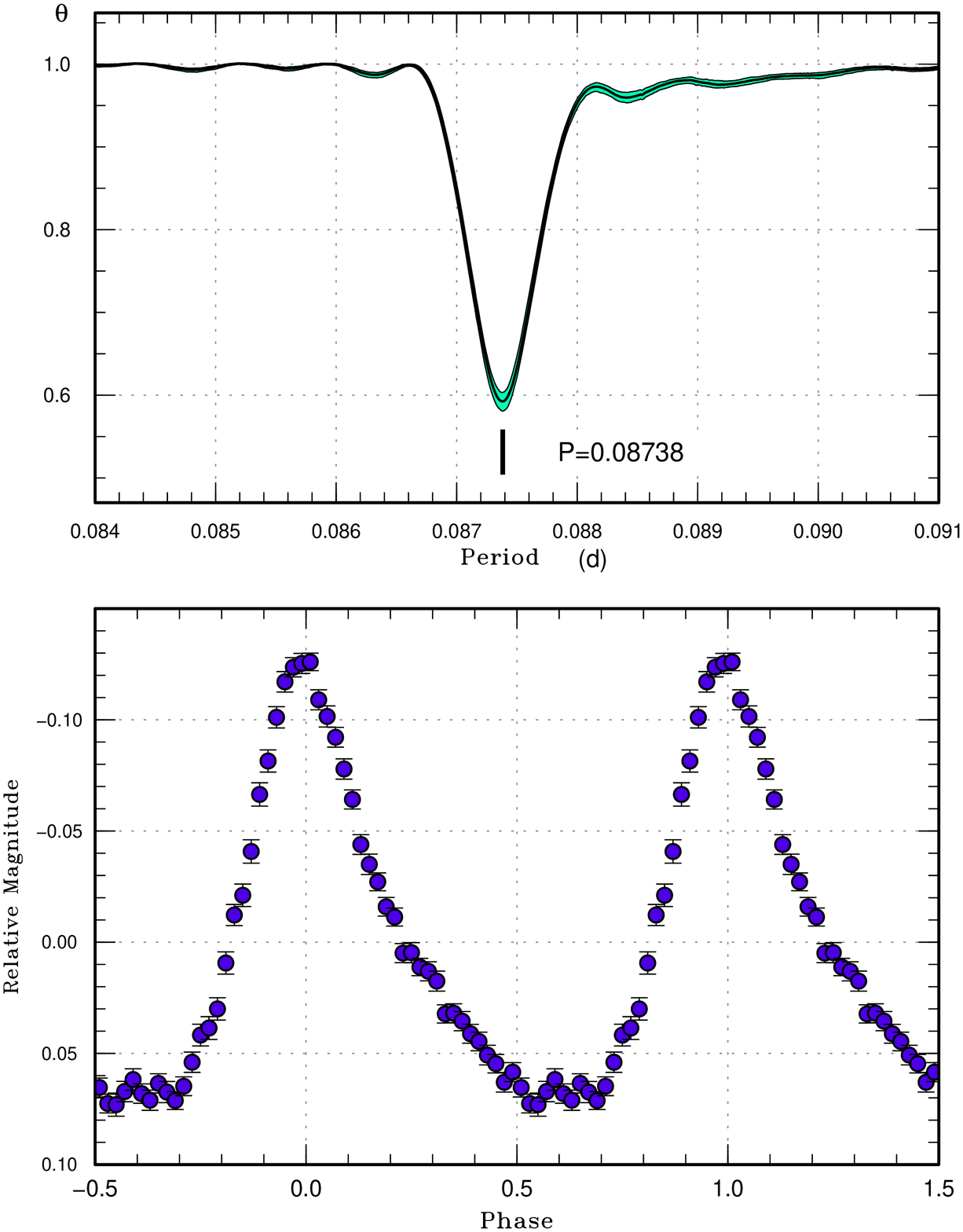}
  \end{center}
  \caption{Superhumps in V516 Lyr.  The figure is produced as in the
     same way in \citet{Pdot3} for the SC data for BJD 2455783--2455793.
     (Upper:) Phase dispersion minimization (PDM; \cite{PDM}) analysis.
     (Lower:) Phase-averaged profile.}
  \label{fig:v516shpdm}
\end{figure}

\begin{figure}
  \begin{center}
    \FigureFile(88mm,110mm){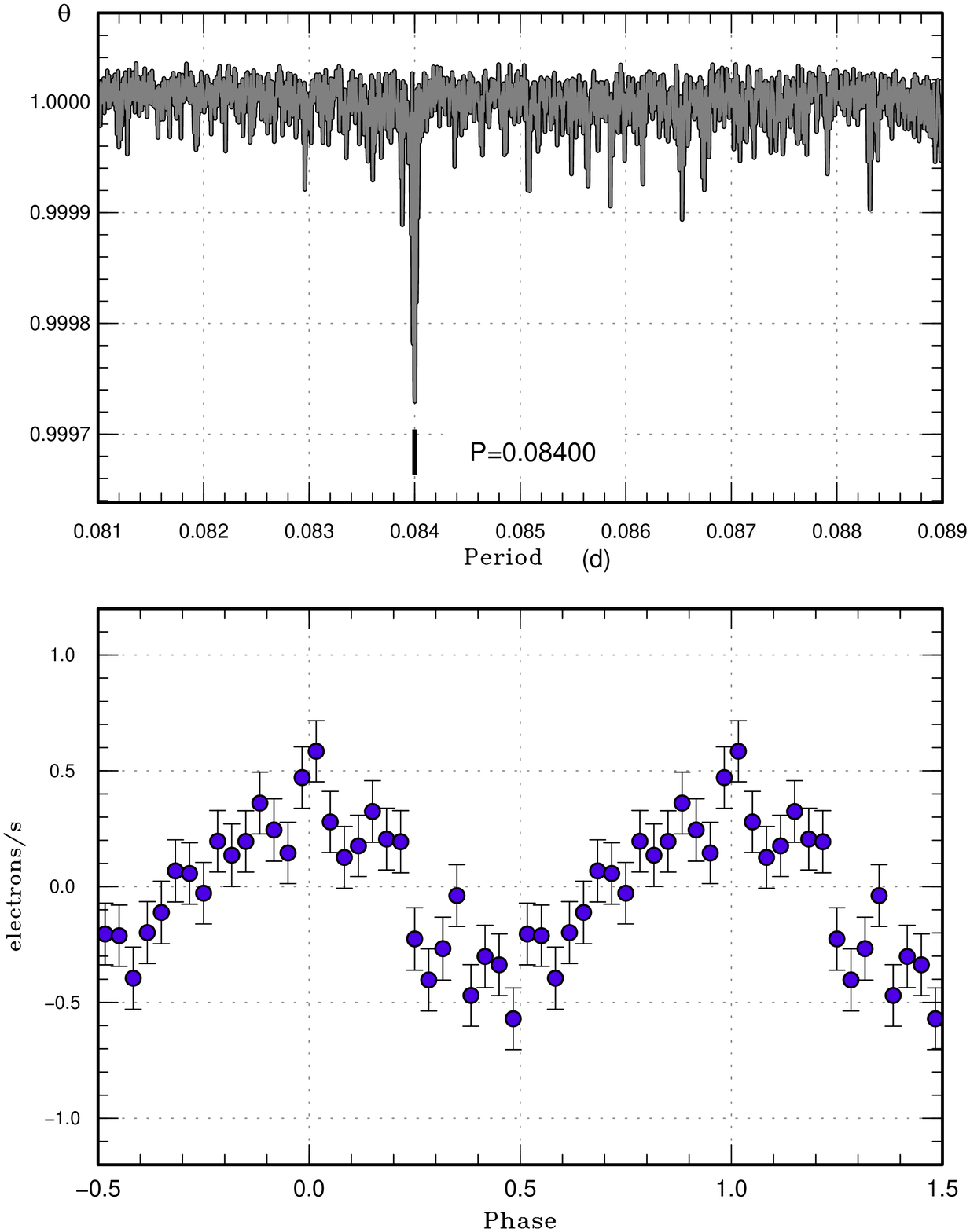}
  \end{center}
  \caption{Orbital signal in V516 Lyr.  The SC data excluding
     two superoutbursts and the interval until the next normal
     outbursts were used.  There were apparently no period other than
     the orbital period.
     (Upper:) PDM analysis.
     (Lower:) Phase-averaged profile.}
  \label{fig:v516porbpdm}
\end{figure}

   The object was observed in Kepler SC in Q8--Q11 and Q14.
Since the object is faint in quiescence ($\sim$30 electrons s$^{-1}$)
and the baseline showed systematic trends, we adjusted the quiescent
level to $Kp=21$ (as in the LC data) by subtracting low-order 
(up to second) polynomials to the observed count rates for each
continuous quarter (duration 67--97~d).  The resultant
quiescent magnitudes are not real, but this treatment is sufficient
since our interest is in the residual pulsed flux.
The mean superhump waveform during one of its superoutburst
(SO No.2 or Q10-5) is shown in figure \ref{fig:v516shpdm}.
We also analyzed the SC data after excluding two superoutbursts and 
the post-superoutburst phase until the next normal outburst
(but not excluding other normal outbursts), i.e. a combined data
before BJD 2455592, BJD 2455612--2455777 and after BJD 2455805
(figure \ref{fig:v516porbpdm}).  The only detected
signal was 0.083999(8)~d, which is most likely the orbital period.
We could not confirm a period of 0.087478~d reported in
\citet{how13KeplerCVs}.
There was no hint of persisting negative superhumps or positive
superhumps in quiescence in this interval.  The fractional
superhump excess is 4.0\%, typical for this orbital period.

\begin{figure*}
  \begin{center}
    \FigureFile(160mm,120mm){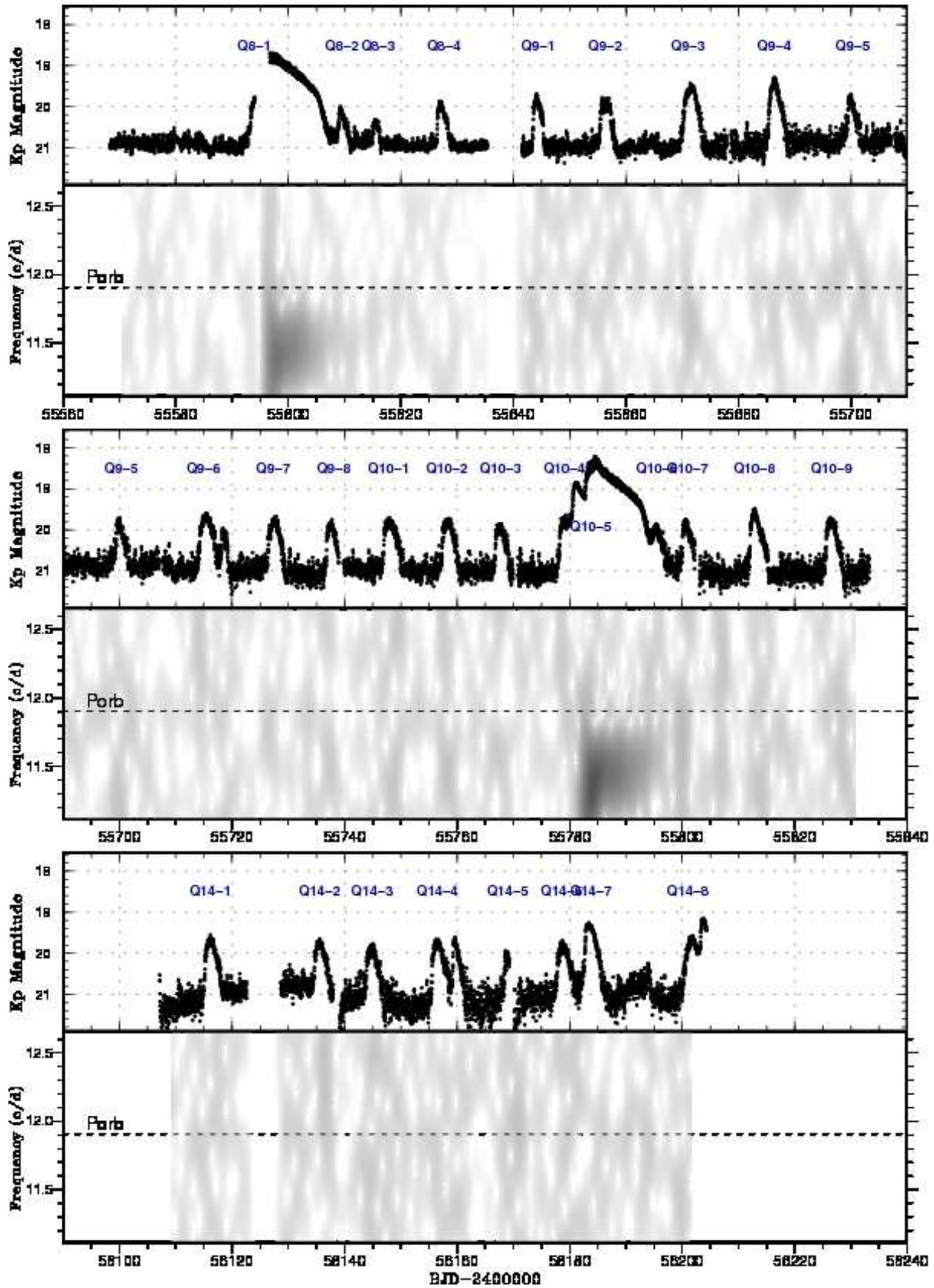}
  \end{center}
  \caption{Two-dimensional Fourier power spectrum of the Kepler
  SC light curve of V516 Lyr.
  (upper:) Light curve; the Kepler data were binned to 0.02~d.
  (lower:) Fourier power spectrum. The width of 
  the moving window and the time step used are 5~d and 0.5~d,
  respectively.}
  \label{fig:v516spec2d}
\end{figure*}

\begin{figure*}
  \begin{center}
    \FigureFile(160mm,120mm){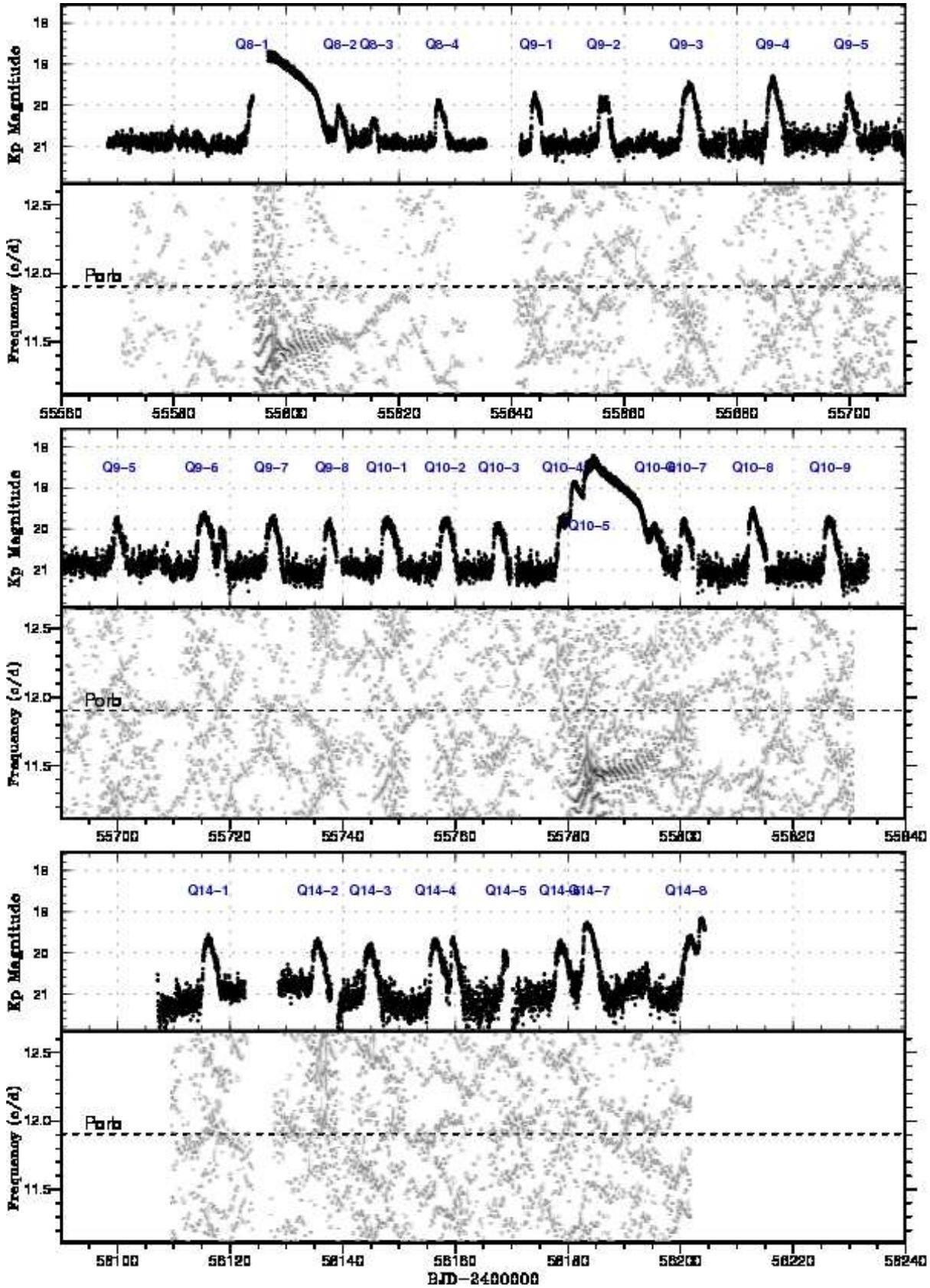}
  \end{center}
  \caption{Two-dimensional Lasso power spectrum of the Kepler
  SC light curve of V516 Lyr.
  (upper:) Light curve; the Kepler data were binned to 0.02~d.
  (lower:) Lasso power spectrum ($\log \lambda=-5.7$). The width of 
  the moving window and the time step used are 5~d and 0.5~d,
  respectively.}
  \label{fig:v516spec2dlasso}
\end{figure*}

   As in V1504 Cyg and V344 Lyr \citep{osa13v344lyrv1504cyg},
we calculated Fourier and Lasso two-dimensional power spectra
(\cite{kat12perlasso}; \cite{kat13j1924})
(figures \ref{fig:v516spec2d}, \ref{fig:v516spec2dlasso}).
Due to the faintness of the object, the power spectra are not
as clear as in V1504 Cyg and V344 Lyr, particularly around quiescence.
During Q8--Q11, the object underwent a relatively regular pattern
of normal outburst and two superoutbursts separated by $\sim$185~d.
The characteristic signals of positive
superhumps during superoutbursts showed an increase of the frequency
(shortening of the superhump period) as reported by
\citet{gar11v516lyratel3507}.  The orbital signal was also sometimes
detected in the Lasso power spectra, but was not clearly seen
in the Fourier power spectra.

   Although there were no clear
persistent negative superhumps as observed in V344 Lyr and V1504 Cyg,
there were several occasions of impulsive (failed) negative superhumps,
which are negative superhumps transiently seen during some normal
outbursts [terminology introduced in \citet{osa13v344lyrv1504cyg};
see also the description of the phenomenon in \citet{woo11v344lyr}].
These impulsive negative superhumps occurred around BJD 2455735 
(outburst Q9-8) and 2455750 (Q10-1), four and three cycles 
before a superoutburst.

\subsection{Superoutburst with Double Precursor}\label{sec:v516lyrdoubleprec}

   The most notable feature in V516 Lyr is a ``double precursor'' in 
SO No. 2, i.e. Q10-4.  Here we discuss two different possibilities 
for the origin of this double precursor.

   The one of possible explanations is an explanation same 
as that of the double outburst presented in \ref{sec:v516lyrglobal}.  
As already discussed there, the preceding outburst 
is an inside-out outburst in which the heating wave 
is reflected in the middle of the disk. As discussed in subsection
3.2 in \citet{osa13v1504cygKepler}, this type of outburst is not accompanied 
with the disk expansion and so it does not trigger the 3:1 resonance tidal 
instability. On the other hand, since the following outburst is an 
outside-in outburst, it is accompanied with the disk expansion and thus 
it can trigger the 3:1 eccentric tidal instability and it can start 
superhumps and a superoutburst. 

\begin{figure}
  \begin{center}
    \FigureFile(88mm,70mm){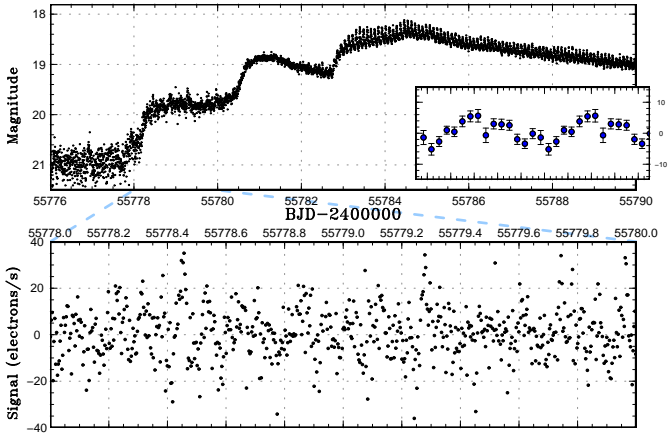}
  \end{center}
  \caption{Possible Impulsive negative superhumps during the double precursor.
     (Upper:) Kepler SC light curve (binned to 0.003~d).
     (Middle:) Pulsed flux (binned to 0.006~d).
     The inset on right edge the upper panel is the profile of
     impulsive negative superhumps averaged for BJD 2455778.2--2455780.0
     (period=0.0828~d).
     }
  \label{fig:v516doubleprec}
\end{figure}

   Another possibility is that related to the failed precursor 
due to an appearance of impulsive negative superhumps.  
During the preceding precursor (BJD 2455778),
negative superhumps seemed to have appeared [period for
BJD 2455778--2445780 was 0.0828(3)~d with the PDM method;
figure \ref{fig:v516doubleprec}]. 
However, this signal diminished during the following precursor 
and instead positive superhumps appeared just as in
ordinary precursor outbursts of V344 Lyr and V1504 Cyg.
In this interpretation, the preceding precursor outburst triggered
the development of impulsive negative superhumps which
were unable to sustain the disk in a hot state, while the second
precursor successfully triggered the development of eccentric
instability to produce positive superhumps and succeeded in
starting a superoutburst.  This kind of ``failed superoutburst''
due to the impulsive negative superhumps was also seen
in V1504 Cyg (section 4.5.2 and figure 77 in \cite{Pdot3}).
It is interesting to note that this failed superoutburst in V1504 Cyg
occurred in the second normal outburst prior to  a 
following superoutburst
[cf. subsection 2.7 in \citet{osa13v344lyrv1504cyg}].

   The two explanations presented above seem to contradict with each other 
because the inside-out type precursor with Smak's type Bb 
in the first explanation is not expected to excite impulsive negative 
superhumps. This is  because 
the disk does not expand with this type of outburst.  
Although this may contradict with observed impulsive negative superhumps, 
we would like to keep both explanations
as possibility at this moment, since the amplitudes of the seeming
impulsive negative superhumps were close to the detection
limit and it is difficult to make a definite conclusion.

We note, however, there have been at least two objects
which showed inside-out-type rise in the superoutburst:
ER UMa \citep{ohs12eruma} and BZ UMa (fig 150 in \cite{Pdot}).
In figure 2 in \citet{ohs12eruma} for ER UMa, the transition 
from negative to positive
superhumps took place $\sim$2~d after the start of
the superoutburst with a kink in the light curve. This case causes 
no problem as an expansion of the disk occurs if the inside-out outburst is 
Smak's type Ba in which the heating front propagating outward reaches the 
outer edge of the disk.
In BZ UMa, there appears to be a weak second precursor at 
around BJD 2454203, after which superhumps developed, 
although a double-nature of the precursor is not as evident as in V516 Lyr.
This superoutburst appeared to be understood as an inside-out-type
precursor followed by an outside-in type second precursor,
which were not well resolved from the initial precursor.

\subsection{$O-C$ Analysis of Superhumps}\label{sec:v516oc}

\begin{figure*}
  \begin{center}
    \FigureFile(160mm,230mm){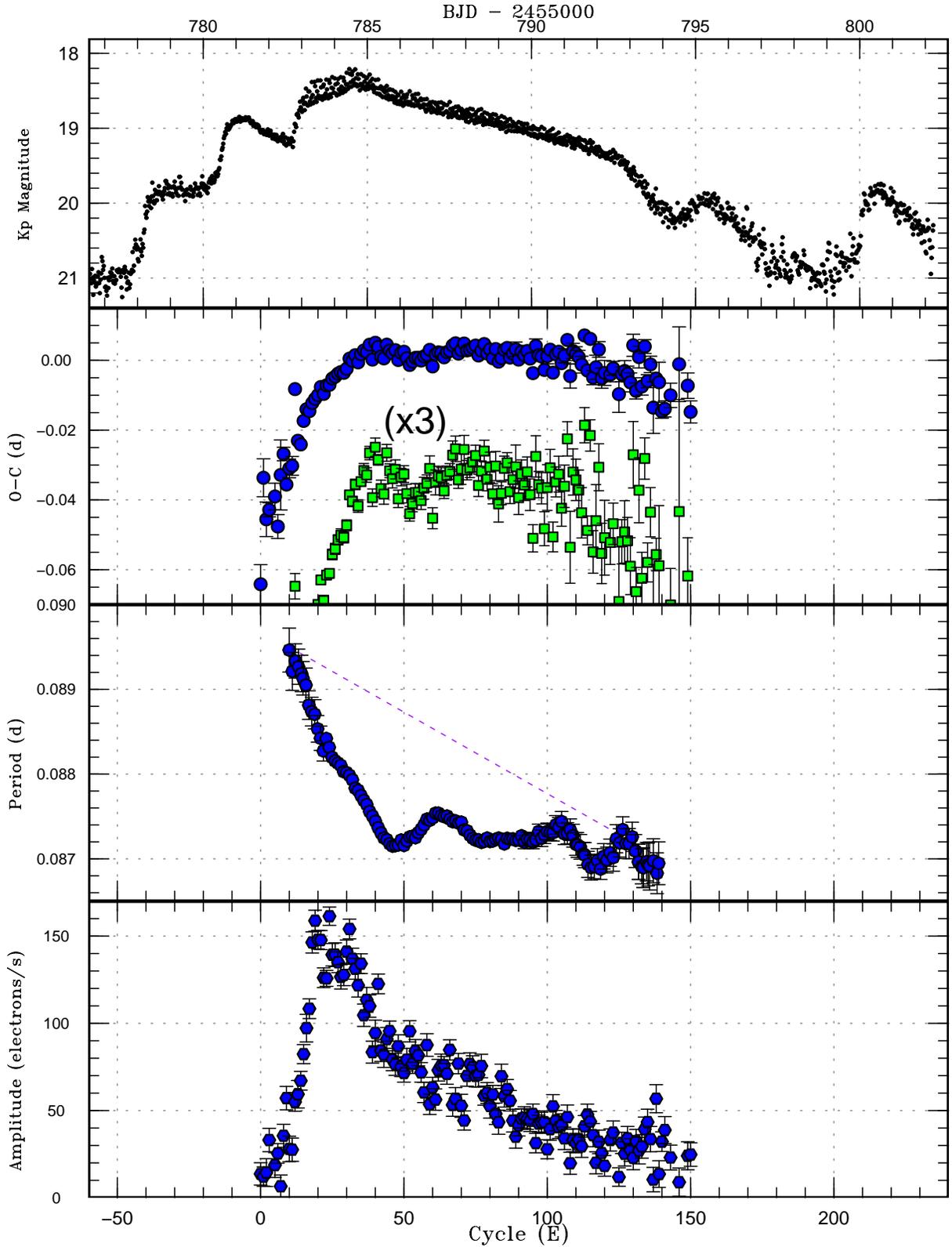}
  \end{center}
  \caption{Analysis of the superoutburst (BJD 2455780--2455797)
  in V516 Lyr.
  From top to bottom:
  (1): Light curve.  The quiescent flux level was artificially
  adjusted to $Kp=21$ level.  Note that the magnitudes in the faintest
  part are not reliable.
  (2): $O-C$ diagram of the superhumps filled circles).
  Filled squares represent $O-C$ values multiplied by three (and shifted
  arbitrarily by a constant) to better visualize the subtle variation
  during the superoutburst.  A ephemeris of
  BJD $2455781.750+0.08732 E$ was used to draw this figure.
  (3): Period determined from the $O-C$ diagram.
  The period was determined by a linear regression to 
  the adjacent 20 times of maxima ($\sim$1.5~d).
  The window width is indicated by a horizontal bar at
  the upper right corner.
  The error refers to 1$\sigma$ error in this regression.
  Note that this process introduce artificial smoothing of
  the period variation and the error is only a random error
  which is likely smaller than the actual error.
  The dashed line is an imaginary global trend of the period 
  decrease in the absence of the pressure effect
  assuming that the period changes linearly
  (will be discussed in subsection \ref{sec:globaldecrease}).
  (4): Amplitudes of the superhumps.
  }
  \label{fig:v516humpall}
\end{figure*}

\begin{figure}
  \begin{center}
    \FigureFile(88mm,110mm){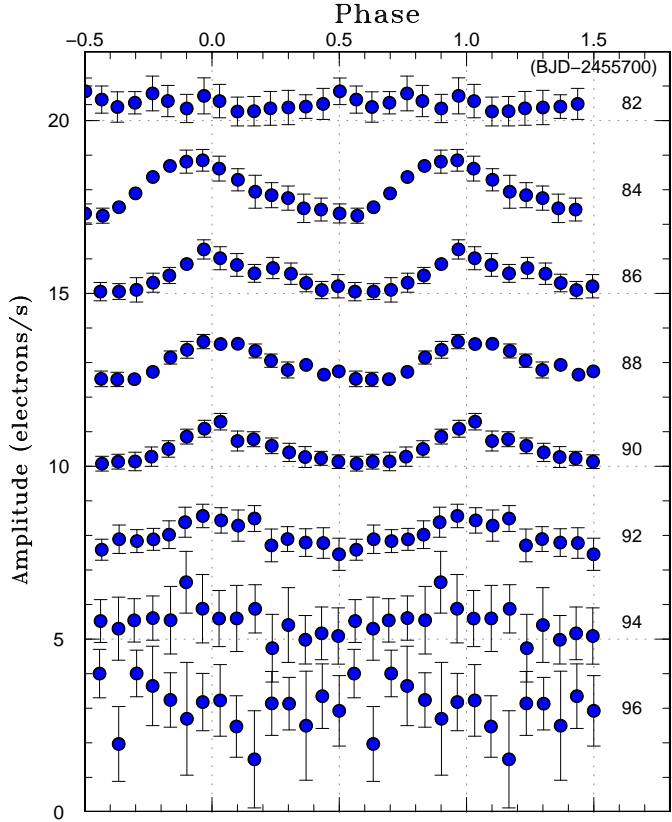}
  \end{center}
  \caption{Variation of superhump profiles of V516 Lyr.
  The phase-averaged profiles were calculated for 2-d segments.
  The phases were calculated using the same ephemeris as in
  figure \ref{fig:v516humpall}.
  }
  \label{fig:v516prof}
\end{figure}

   We calculated the $O-C$ diagram of the superoutburst around
BJD 2455780--2455797 (the superoutburst with double precursors).
The $O-C$ diagram was roughly composed of stages A--B--C, as
described in \citet{Pdot}.  The transitions from stage A to B
and stage B to C were rather smooth compared to short-$P_{\rm orb}$
systems described in \citet{Pdot}, and more resembles the
long-$P_{\rm orb}$, high-$\dot{M}$ system V344 Lyr \citep{Pdot3}.
Although it is not very clear what segments in $E$ are most
adequate to determine the period in each stage, we adopted
periods of 0.08732(3)~d and 0.08725(3)~d for stage B
($38 \le E \le 70$) and stage C ($70 \le E \le 120$), respectively.
During stage B, a period increase at
$P_{\rm dot} = +27(7) \times 10^{-5}$ was recorded.

   There was a break in the $O-C$ diagram around the time
when the amplitude of the superhump reached a maximum.
As in the background dwarf nova of KIC 4378554, we refer to
these stages as A1 ($E \le 15$) and A2 ($16 \le E \le 38$).
The mean periods of superhumps during stages A1 and A2 were
0.0894(4)~d and 0.08814(3)~d, 6.4(5)\% and 4.93(4)\%
longer than the orbital period, respectively.

   The profile of the superhumps (figure \ref{fig:v516prof})
did not show the strong secondary peak as seen in V344 Lyr
(\cite{woo11v344lyr}; \cite{Pdot3}).  There was no clear
transition to traditional late superhumps having
a $\sim$0.5 phase jump.  During the late stage of the superoutburst,
the amplitudes of the superhumps (including the secondary maxima,
if they existed) decreased to an undetectable
limit (amplitude $\lesssim$1 electrons s$^{-1}$).
These amplitudes were no larger than that of the orbital variations
in quiescence (subsection \ref{sec:v516lyrsc}).

\subsection{LC Data}

\begin{figure*}
  \begin{center}
    \FigureFile(110mm,110mm){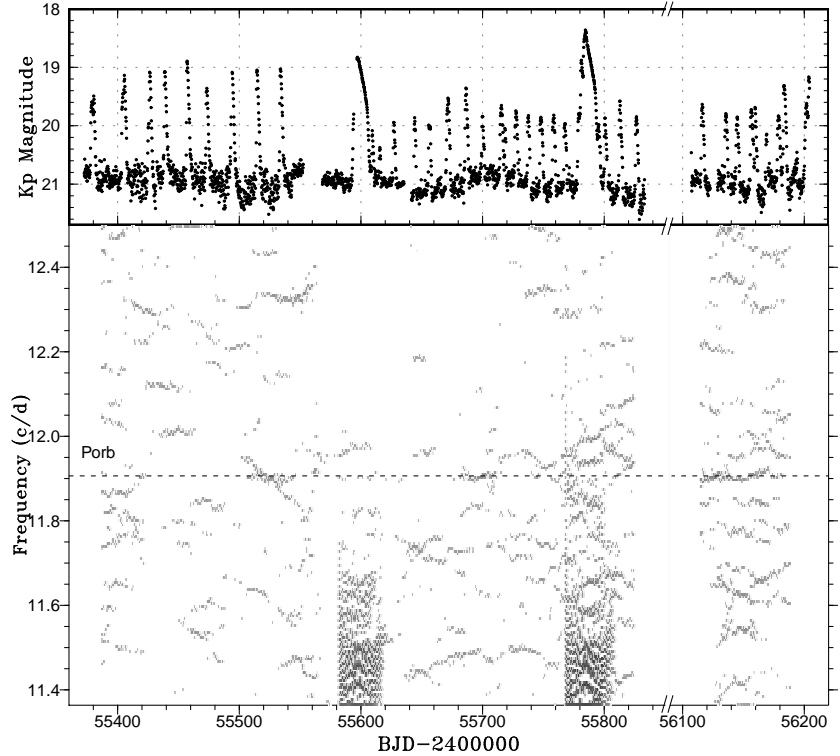}
  \end{center}
  \caption{Two-dimensional Lasso power spectrum of the Kepler
  LC light curve of V516 Lyr.
  (Upper:) Light curve; the Kepler data were binned to 0.02~d.
  Since there is no stable long-term zero point in the Kepler data,
  we artificially added a constant to the flux of each quarter to obtain
  the similar level (mag $\sim$21) in quiescence.
  (Lower:) Lasso power spectrum ($\log \lambda=-5.75$). The width of 
  the moving window and the time step used are 30~d and 1~d,
  respectively.}
  \label{fig:v516lcspec2dlasso}
\end{figure*}

   We also made a two-dimensional Lasso power spectrum of the LC
data (figure \ref{fig:v516lcspec2dlasso}).  Since the number of
data points was much smaller than in the SC data, we could only
detect relatively persistent signals.  The faintness of V516 Lyr
also made the analysis more difficult than in V344 Lyr or V1504 Cyg
\citep{osa13v344lyrv1504cyg}.  We adopted a window of 30~d, which
was shown to give the best signal-to-noise for the orbital modulation,
and hence is expected to detect signals with similar strengths.
This length of the window was too long to resolve any variation
of positive superhumps during the superoutbursts, and these
positive superhumps were only detected as broad bands.
In figure \ref{fig:v516lcspec2dlasso}, however, there seem to be
transient weak signals of possible negative superhumps around
the frequencies 12.32--12.35 c/d in segments BJD 2455500--2455560
(this may also be extended to a possible frequency 12.30 c/d 
around BJD 2455420--2455450),
2455730--2455760 and 2456120--2456150.  Although the reality of
these signals is difficult to confirm, we suspect that they were
likely present because the initial segment corresponds to 
a period of decreased frequency of outbursts (type ``L'' in
\cite{osa13v1504cygKepler}, and the second segment corresponds
to a period of impulsive negative superhumps discussed in
subsection \ref{sec:v516lyrsc}.  The $\epsilon^*$ for these
possible negative superhumps was 3.5--3.7\%.

   Although the result is not very conclusive, the tendency
that negative superhumps suppress the occurrence of outbursts
(\cite{ohs12eruma}; \cite{osa13v1504cygKepler}; \cite{zem13eruma})
seems to be valid for V516 Lyr.  If the signals around
BJD 2455420-2455450 were also indeed of negative superhumps,
there was an increase of the frequency of negative superhumps
as the phase of the supercycle progresses, which has been confirmed
in V1504 Cyg and V344 Lyr (\cite{osa13v1504cygKepler};
\cite{osa13v344lyrv1504cyg}).  We expect future Kepler SC runs
(if available) during a state with negative superhumps will
provide us better characterization of the phenomena.
The Kepler results up to now appears to suggest that 
negative superhumps in quiescent SU UMa-type dwarf novae are
more prevalent that have been considered.

   An additional feature in the two-dimensional power spectrum
analysis is that the orbital signal was more strongly detected in Q14
than in other quarters.  Such a variation of the strength of
the orbital signal was also seen in V1504 Cyg and V344 Lyr
(\cite{osa13v1504cygKepler}; \cite{osa13v344lyrv1504cyg}).

\subsection{Interpretation of Historical Data}

   Although \citet{moc03v516lyr} suggested the SS Cyg-type
classification, it has now become evident that this object
is an SU UMa-type dwarf nova.  We reexamined the material in
\citet{moc03v516lyr}.  The phased light curve (figure 6)
in \citet{moc03v516lyr} indicated a mean period of 17.7298~d, 
which appears to be consistent with the Kepler observations. 
Their long-term light curve
(figure 6, object B8) showed at least six major outbursts
most of which reached maxima of $V=19$.  One of them
(in their window No. 12) lasted at least 5~d, and this outburst
must have been a superoutburst.  On one occasion (in their window
No. 3), the object was detected as faint as $V=22\sim23$, more
than 1 mag fainter than its ordinary quiescence.  Since this
object is not eclipsing, this phenomenon may have been caused
by a temporary reduction of the mass-transfer rate.
It is difficult to check whether a similar phenomenon was 
recorded in the Kepler data due to the faintness of the object
and the highly variable zero-point.

\section{Discussion}

\subsection{Appearance of Superhumps and its Implication to
the Thermal-Tidal Instability Theory}

   In the background dwarf nova of KIC 4378554 and V516 Lyr,
all superoutbursts took a form of the precursor--superoutburst
combination, same as reported in V1504 Cyg and V344 Lyr
(\cite{sti10v344lyr}; \cite{woo11v344lyr}; \cite{can10v344lyr};
\cite{can12v344lyr}; \cite{osa13v1504cygKepler}; 
\cite{osa13v344lyrv1504cyg}).
In both systems, superhumps always started to appear during
the fading branch of precursor and reached the maximum
amplitude around the peak of the main superoutburst.
This sequence of a precursor --- development of superhumps ---
main superoutburst very well fits the picture of the TTI
model (\cite{osa89suuma}; \cite{osa13v1504cygKepler}; 
\cite{osa13v344lyrv1504cyg}) and strengthens the universal
application of the TTI theory to two more additional SU UMa-type
dwarf novae.

   In the case of V585 Lyr, the superhump took $\sim$7~d to reach
the maximum amplitude.  This long waiting time compared to 
the background dwarf nova of KIC 4378554 and V516 Lyr
can be naturally understood as an effect of a lower $q$ in V585 Lyr.
Since the growth rate of the eccentric mode is proportional
to $q^2$ \citep{lub91SHa}, a typical $q=0.10$ for the short period
V585 Lyr requires $\sim$3 times longer time than in
a case of $q=0.18$ in V516 Lyr.  Since it took $\sim$2--3~d to
reach the maximum amplitude in V516 Lyr and the background 
dwarf nova of KIC 4378554, the agreement with this expectation
is very good and reinforces the TTI theory.

\subsection{Period Variation of Superhumps}

\subsubsection{Global Variation}\label{sec:globaldecrease}

   We studied the period variation of superhumps in all three SU UMa stars 
by using the $O-C$ data.
We did not use period finding methods such as Fourier transform or 
PDM because the $O-C$ data are more sensitive for detecting
the period variation and this method gives a more stable result
when the sampling rate is very low.  The results are shown in
the third panel of figures \ref{fig:nik1humpall} [which is
an improvement of \citet{bar12j1939}], \ref{fig:v585humpall}, and 
\ref{fig:v516humpall}.  We note, however, there appeared some 
wiggles in V585 Lyr, which should not be considered as real
since the LC sampling rate in this object was almost exactly one 
third of the superhump period, which resulted in artificial signals.
All three objects show a global decrease in
the superhump period (shown by a broken line in the figures):
from 0.0778~d to 0.0765~d in the background dwarf nova 
of KIC 4378554, from 0.0613~d to 0.0600~d in V585 Lyr and
from 0.0895~d to 0.0870~d in V516 Lyr. The global decrease in the superhump 
period during superoutburst in these three stars are thought to be produced 
by monotonic decrease in the disk radius (\cite{osa13v344lyrv1504cyg}). 
  
   Among them, V516 Lyr has a photometric orbital period ($P_{\rm orb}$) 
and we can determine the precession rate of the eccentric mode (the superhump 
mode) in the accretion disk.  
As discussed in \citet{osa13v344lyrv1504cyg}, it is very convenient 
to introduce the fractional superhump deficiency in the frequency unit 
$\epsilon^* \equiv (\nu_{\rm orb}-\nu_{\rm SH})/\nu_{\rm orb}
=1-P_{\rm orb}/P_{\rm SH}$ where $\nu_{\rm orb}$ and $\nu_{\rm SH}$ 
are orbital and superhump frequencies, respectively. The ordinary superhump 
period excess, $\epsilon \equiv (P_{\rm SH}-P_{\rm orb})/P_{\rm orb}$
is related to this quantity by $\epsilon^*=\epsilon/(1+\epsilon)$. 
The quantity, $\epsilon^*_{+}$, is then related to the apsidal precession 
rate of the eccentric disk, $\nu_{\rm prec}$, 
by $\epsilon^*_{+}=\nu_{\rm prec}/\nu_{\rm orb}$. 
The quantity $\epsilon^*_{+}$ in the growing stage of positive superhumps is 
thought to reflect the precession rate at the 3:1 resonance, 
because the superhump wave is restricted to the resonance region 
(\cite{osa13v344lyrv1504cyg}).
The precession rate in this condition can be expressed as
(\cite{hir90SHexcess}; \cite{hir93SHperiod}):
\begin{equation}\label{equ:presfreq}
\epsilon^*_{+}=
\frac{\nu_{\rm prec}}{\nu_{\rm orb}}=\frac{q}{\sqrt{1+q}}
\Bigl[\frac{1}{2}\frac{1}{\sqrt{r}}\frac{d}{dr}\Bigr(r^2\frac{dB_0}{dr}\Bigr)\Bigr],
\end{equation}
where  $r$ is the radius from the white dwarf primary 
(in unit of the binary separation).
$B_0$ is written as 
\begin{equation}\label{equ:laplace}
B_0(r)=b_{1/2}^{(0)}/2=\frac{2}{\pi}\int_0^{\pi/2}\frac{d\phi}{\sqrt{1-r^2\sin^2 \phi}},
\end{equation} 
which is the Laplace coefficient of the order 0 in celestial mechanics
\citep{sma53Celestialmechanic}.
Using $\epsilon^*=0.060(4)$ for stage A1 superhumps, we obtained
a mass ratio of $q=0.18(2)$ for V516 Lyr where we used the radius 
of the 3:1 resonance by $r_{3:1}=3^{(-2/3)}(1+q)^{-1/3}$.

The global decrease in the superhump periods and hence in the quantity 
$\epsilon^*$ indicates that the disk radius decreased 
during superoutbursts in these three SU UMa stars. 
The $\epsilon^*$ around the end of the plateau phase of 
V516 Lyr was 3.4\%.
This value corresponds to a radius of 0.36$A$ if we ignore
the pressure effects, where $A$ is the binary separation. 

\subsubsection{Pressure Effects}

   As discussed in \citet{osa13v344lyrv1504cyg}, we consider that
the temporary deviation from the general trend of the period variation
is caused by the pressure effects.
The pressure effects bring about a decrease in the superhump eigenfrequency
\citep{hir93SHperiod} and thus a decrease in the (positive) 
superhump period \citep{osa13v344lyrv1504cyg}, 
besides the effect of disk radius variation. 
We interpret that the temporary deviation from the general trend
(temporary decrease in period) commonly seen in three systems was
caused by the pressure effects.  Although the strength of this effect 
seems to be different from object to object,
the effect is the strongest around the early phase of the superoutburst,
particularly around the stage A--B transition.  This is 
because the disk temperature is the highest
near the start of the superoutburst.  

   A much stronger downward 
deviation seen in V585 Lyr in a period of BJD 2455235--2455239,
may be understood as a stronger pressure effect. 
As discussed in \citet{osa13v344lyrv1504cyg}, 
the precession rate of the eccentric mode is given by two 
different terms of the dynamical prograde precession by 
the tidal perturbation of the secondary star and of 
the retrograde precession by pressure effects. 
V585 Lyr is thought to have a lowest $q$ among the three 
SU UMa stars discussed here judging from their superhump periods. 
The pressure effects may affect the precession frequency 
of the eccentric mode (and the superhump period) more strongly 
in a low-$q$ system such as V585 Lyr.  This is because 
the prograde precession rate by the tidal effect is lower 
in a lower-$q$ system while the pressure effects work for 
retrograde precession similarly regardless of $q$.
Thus the relative importance of pressure effects may increase 
if we go to a system of lower $q$. 

\subsubsection{Period after Superoutburst}

   After the superoutburst, the period tends to remain at
a short value, probably reflecting a smaller disk radius.
It looks like that the period increased around the rebrightening
in the background dwarf nova of KIC 4378554 and V585 Lyr
which may reflect the increase in the disk radius due to the thermal
instability (cf. \cite{osa13v344lyrv1504cyg}).  The periods
of the superhump around this time, however, may not be very 
reliable because the amplitudes of superhumps were very low.
The period variation around the time of ``mini-rebrightenings''
of V585 Lyr was not very informative due to the noisy $O-C$ diagram 
and the time-resolution was insufficient 
to resolve ``mini-rebrightenings''.

\subsection{Superoutburst without a Precursor}

   The outburst of V585 Lyr recorded with Kepler did not show 
a precursor outburst.   This is currently the only known 
superoutburst without a precursor in the Kepler data.
In the framework of the thermal-tidal instability model \citep{osa05DImodel},
this superoutburst can be understood as follows: if the mass is 
sufficiently accumulated in the accretion disk during a long quiescence, 
the next normal outburst could bring the disk to expand 
to the tidal truncation radius, by passing the 3:1 resonance. The expansion 
of the disk is stopped at this radius and the disk is then 
in fully hot state, and the viscous
depletion of matter ensues even when the eccentric tidal instability
has not yet developed. This allows the disk to develop a growth of 
eccentric structure (and thus superhumps) a few days 
after the outburst maximum. This situation corresponds to ``case B''
superoutburst discussed in \citet{osa03DNoutburst} and \citet{osa05DImodel}.  

As shown by
\citet{kry01v585lyrv587lyr}, V585 Lyr also showed a superoutburst
with a precursor [``case A'' in \citet{osa03DNoutburst}].
This indicates that the properties of V585 Lyr are not so extreme
as WZ Sge-type dwarf novae, which predominantly show superoutbursts 
without precursor 
and very rarely\footnote{
  \citet{ric92wzsgedip} reported a possible dip in the early
  stage of the outburst of DV Dra in 1991.  The object underwent
  a superoutburst in 2005 \citep{Pdot} and its property was
  that of a WZ Sge-type dwarf nova ($P_{\rm orb}$=0.0588~d).
  V585 Lyr and DV Dra may be similar objects.
  Recently, a dwarf nova with a very
  large outburst amplitude ($\sim$8 mag), OT J075418.7$+$381225
  = CSS130131:075419$+$381225, showed a precursor outburst
  (vsnet-alert 15438).  This object has a long superhump period
  of 0.07194~d (vsnet-alert 15438) and may be different from
  an ordinary WZ Sge-type dwarf nova.
} show  a precursor-superoutburst-type outburst.
[For a recent review of WZ Sge-type dwarf novae, see
\cite{kat01hvvir}; \cite{Pdot}].
The Kepler data clearly demonstrate that the same object can
show two types of superoutbursts, with or without a precursor.

\subsection{Late-Stage Superhumps}

   It has been well known that the prototypical Kepler SU UMa-type
dwarf novae V344 Lyr and V1504 Cyg showed the development of
secondary humps which evolved into the strongest signal after
the fading from the superoutburst \citep{woo11v344lyr}.
This picture agrees with the traditional interpretation of
the late superhumps \citep{vog83lateSH}, i.e. the varying 
energy release on an eccentric disk around the stream impact point.
The degree of the development of the secondary hump is,
however, reported to be different between V344 Lyr and V1504 Cyg,
and the latter has much less prominent secondary humps \citep{Pdot3}.

   In the present study, none of three objects showed neither
a strong sign of the secondary hump nor a $\sim$0.5 phase jump
expected for traditional late superhumps.  This result seems
to confirm the ground-based results (\cite{Pdot}; \cite{Pdot2};
\cite{Pdot3}; \cite{Pdot4}) that most of SU UMa-type dwarf novae,
particularly low-$\dot{M}$ ones, do not show a $\sim$0.5 phase jump
characteristic to traditional late superhumps.  From the low
amplitudes of post-superoutburst superhumps in the three objects,
we may conclude that there was no evidence of enhanced mass transfer
during the superoutburst.

   It appears that V344 Lyr is rather exceptional in its strength of
the secondary hump.  This may be related to its high $\dot{M}$
as inferred from its short supercycle.

   The present result for V585 Lyr clearly confirms that
so-called stage C superhumps survive after the fading
from the superoutburst without showing a $\sim$0.5 phase jump.
This has demonstrated our suggestion that at least some
``late superhumps'' (supposing a $\sim$0.5 phase jump)
can be understood as a result of mistaken cycle count for
stage C superhumps with a shorter period \citep{Pdot}.

\section{Conclusion}

   We studied Kepler SC and LC light curves of three SU UMa-type
dwarf novae: the background dwarf nova of KIC 4378554, V585 Lyr, 
and V516 Lyr.

Both the background dwarf nova of KIC 4378554 and V516 Lyr showed 
a similar combination of precursor-main superoutburst, during 
which superhumps always started to appear in the fading 
branch of the precursor (the ``type A'' superoutburst 
in \cite{osa03DNoutburst} and \cite{osa05DImodel}).  
This behavior is common to V344 Lyr and V1504 Cyg 
and strongly supports the TTI theory for the origin of the superoutburst.

   V585 Lyr is a borderline object between SU UMa-type dwarf novae and
WZ Sge-type dwarf novae.  The Kepler LC data recorded one superoutburst
without a precursor.  The development of this outburst can be
understood within the TTI model as a case in which the disk with
mass sufficiently accumulated first expanded to reach the tidal 
truncation radius by a triggering normal outburst and then 
the tidal instability and the superhump developed 
after the maximum (the ``case B'' 
in \cite{osa03DNoutburst} and \cite{osa05DImodel}).  
The subsequent development of
the $O-C$ diagram perfectly confirmed the existence of three 
distinct stages (A--B--C) for short-$P_{\rm orb}$ SU UMa-type
dwarf novae inferred from ground-based observations.
This observation made the first clear Kepler detection of
the positive period derivative commonly seen in the stage B
superhumps in such dwarf novae.
There were recurring ``mini-rebrightenings'' between the superoutburst
and the rebrightening.

   V516 Lyr is an SU UMa star similar to V344 Lyr but the frequencies of 
both normal outburst and superoutburst are lower than those of V344 Lyr, 
which suggests its lower mass-transfer rate than that of V344 Lyr. 
One of the most interesting aspects in the Kepler light curve of V516 Lyr 
is the appearance of double outbursts. In the double outburst, the preceding 
outburst is of the inside-out nature while the following outburst is of 
the outside-in nature.  The double outbursts tend to occur in groups.
We have determined the orbital period
of V516 Lyr to be 0.083999(8)~d.  One of superoutbursts of
V516 Lyr was preceded by a double precursor.  The 
preceding precursor failed to trigger a superoutburst and the following 
precursor triggered a superoutburst by developing positive superhumps.
There were possible negative superhumps
in the LC data of V516 Lyr.  One occasion with an appearance of 
possible negative superhumps coincided with that of
a reduced number of normal outbursts,
strengthening the conclusion for V1504 Cyg and V344 Lyr 
(\cite{osa13v1504cygKepler}; \cite{osa13v344lyrv1504cyg}) that
the negative superhumps tend to suppress normal outbursts.   

   None of these three SU UMa-type
showed the strong signature of a transition to the dominating
stream impact-type component of superhumps.  This finding suggests
that the case of V344 Lyr is rather exceptional, which is probably
associated with a high $\dot{M}$.  There was no strong
indication of enhanced mass-transfer following the superoutburst.

   We have also developed a method to analyze superhumps of 
short-$P_{\rm SH}$ SU UMa-type dwarf novae in the Kepler LC data by 
modeling the observation and applying the MCMC method.  Since most of 
the new dwarf novae included in the Kepler field by chance 
(cf. \cite{bar12j1939}) are expected to be observed only with LC runs,
and a large fraction of dwarf novae is expected to be short-$P_{\rm SH}$
systems \citep{gan09SDSSCVs}, our method will be effective in
characterizing these dwarf novae. 

\medskip

This work was supported by the Grant-in-Aid for the Global COE Program
``The Next Generation of Physics, Spun from Universality and Emergence"
from the Ministry of Education, Culture, Sports, Science and Technology
(MEXT) of Japan.
We thank the Kepler Mission team and the data calibration engineers for
making Kepler data available to the public.

\appendix

\section{Method to Handle Kepler Pixel Image Data with R}
\label{sec:handlepixel}

   \citet{bar12j1939} introduced {\tt PyKE} routines to extract
custom apertures.  We introduce a different convenient way:
by using {\tt FITSio} package of {\tt R}, and loading the target
pixel FITS file in a variable {\tt d}, one can easily obtain
a time-series data for pixel and draw pixel light curves
as in figure 2 in \citet{bar12j1939} and make custom-aperture
photometry without a special library.  To load a FITS pixel data,
we can use the following command (the file name was split due
to its long length; it should be written in a single line
in actual execution):
\begin{verbatim}
library(FITSio)
d <- readFITS("kplr004378554-2009350155506
     _lpd-targ.fits")
\end{verbatim}
The resultant \verb|d$col| contains the table values in 
the FITS file.  The counts for individual pixel number \verb|n|
are stored in \verb|d$col[[5]][,n]|, where 5 means the fifth
member of the FITS data (pixel counts).  For example,
we can draw a light curve for pixel \verb|n|=3 by:
\begin{verbatim}
plot(d$col[[5]][,3])
\end{verbatim}
If one wish to add three pixels \verb|n|=3, 4, 8, we can
get the summed values by:
\begin{verbatim}
f <- d$col[[5]][,3] + d$col[[5]][,4] +
     d$col[[5]][,8]
\end{verbatim}
and
\begin{verbatim}
plot(f)
\end{verbatim}
to draw the light curve for the custom aperture.

\section{Reconstruction of the Superhump Profile using Kepler LC Data}
\label{sec:recon}

   Since the Kepler LC data were too sparse for a short-period object
like V585 Lyr (only three measurements in one superhump),
we modeled the observation from the template superhump light curve
and obtained the best model referring to the actual observations
(subsection \ref{sec:v585prof}).

   In reconstructing the superhump profile, we subdivided observations 
to 0.5~d segments ($\sim$24 observations).
They were first subtracted for the longer trends caused 
by outbursts, and the mean residuals were adjusted to be zero.
We assumed a template light curve consisting of 20 phase bins.
We can then model the observation by numerically 
integrating the spline-interpolated template for the phases corresponding 
to each Kepler LC exposure (29.4 min) assuming the epoch and period.
We can get from the assumed template a set of 
$\{y_{\rm model}(i)\}$, where $i$ represents the Kepler LC exposure.
We then compare $\{y_{\rm model}(i)\}$ with
the actual observations $\{y_{\rm obs}(i)\}$.  In principle,
we can minimize the difference between $\{y_{\rm model}(i)\}$ and
$\{y_{\rm obs}(i)\}$ by changing the assumed template.
The best-fit model, however, gives a highly structured light curve
not resembling that of a superhump.  This was probably caused by the 
``overexpression'' of the random or systematic variations other than 
the superhump variation.

   We therefore estimated the best superhump profile by using
both the residuals and smoothness using the Bayesian framework.
The values $\theta(j)$ ($j=1,\dots,N$), where $N$ is 
the number of phase bins, represent the template light curve.
The notation follows the appendix of \citet{Pdot2}.
Since Bayes' theorem gives
\begin{equation}
Pr(\theta|D) \propto\, Pr(D|\theta)\pi(\theta),
\end{equation}
where $\theta$ is the model parameter (in the present case, the template
light curve) and $D$ is the observation.  $Pr(D|\theta)$ is
the likelihood and $\pi(\theta)$ is the prior.
$L^0 = \log\{Pr(D|\theta)\pi(\theta)\}$ can be expressed as a form of:
\begin{equation}
L^0 = L_{\rm res}^0 + \lambda L_{\rm sm}^0.
\end{equation}
Here,
\begin{equation}\label{equ:residual}
L_{\rm res}^0 = \sum_i \biggl[ -\log(\sqrt{2\pi}\sigma)
-\frac{\{y_{obs}(i)-y_{model}(i)\}^2}{2\sigma^2} \biggr],
\end{equation}
where $\sigma$ is the observation error.
Following a standard technique in Bayesian analysis,
we then express the condition of smoothness of $\theta$ by
introducing a prior function assuming that second order
differences of $\{\theta(j)\}$ follow a normal distribution.
Then,
\begin{equation}
L_{\rm sm}^0 = \sum_j \biggl[ -\log(\sqrt{2\pi}\sigma_s)
-\frac{\{\theta(j-1)-2\theta(j)+\theta(j+1)\}^2}{2\sigma_{\rm s}^2} \biggr],
\end{equation}
where $\theta(0) = \theta(N-1)$ and $\theta(N+1) = \theta(1)$
reflecting the cyclic condition.
$\sigma_{\rm s}$ is the standard deviation of the normal 
distribution.  Since only the difference of $L^0$ is important
in the MCMC steps, we can omit constant terms.
Since $\sigma$ and $\sigma_s$ are constants, we can now write
in a form
\begin{equation}
L =  \frac{1}{2\sigma^2}(L_1 + L_2) \nonumber \\
\end{equation}
\begin{equation}
L_1 = - \sum_i \{y_{obs}(i)-y_{model}(i)\}^2 \nonumber \\
\end{equation}
\begin{equation}
L_2 = - \lambda^{\prime} \sum_j \{\theta(j-1)-2\theta(j)+\theta(j+1)\}^2 \nonumber \\,
\end{equation}
and $\lambda^{\prime}=\lambda(\sigma/\sigma_{\rm s})^2$.
In actual calculation, we estimated the contribution of $L_1$
and $L_2$ to $L$, and expressed $\beta = L_2/L_1$ as the strength
of the prior.  We changed $\beta$ and selected the most likely 
template light curve.  

\begin{figure}
  \begin{center}
    \FigureFile(80mm,80mm){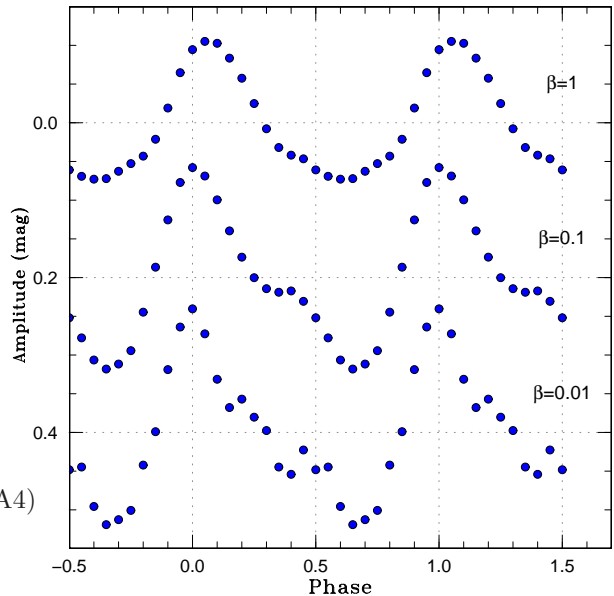}
  \end{center}
  \caption{Dependence on $\beta$ in reconstructing the superhump
  profile from the Kepler LC data.  The data are for V585 Lyr,
  0.5-d segment starting on BJD 2455235 (near the peak amplitude).}
  \label{fig:betadep}
\end{figure}

   By using the actual Kepler data and after comparison with the
superhump light curve taken from the ground-based observation,
we adopted $\beta=0.1$ (for example, $\beta=1$ gives too smooth,
and $\beta=0.01$ gives too structured light curve compared to
the ground-based observation; figure \ref{fig:betadep}).
The result, however, does not very strongly depend on $\beta$.
Since all of the LC measurements
of V585 Lyr have more than 7$\times 10^5$ electrons,
the photon noise is not the dominant source of the error.
We therefore used $\sigma=0.001$ mag for this analysis.
The resultant profile, however, is not sensitive to this value.

   The probability density function (PDF) of model parameters from 
the observed data can be determined by MCMC method (cf. \cite{Pdot2}),
and we adopted the mean and the standard deviation for each phase
of the PDF to obtain the template light curve.  Note that the standard
deviation is not the actual error of the estimate, but only 
a measure of residuals which is strongly affected by the condition 
set by the prior.  A length of 20000 MCMC run was sufficient to
obtain a reasonable PDF and the initial burn-in period was 5000.

   We should note this kind of reverse estimation
may be prone to artificial features, although the solution in the
present case was mostly stable.

\section{Determination of Superhump Maxima in Kepler LC Data}
\label{sec:maxdet}

   In subsection \ref{sec:v585oc}, we determined the times
of superhump maxima by using the MCMC method.  As in subsection
\ref{sec:recon}, we modeled the observation by using a template
with 20 phase bins.  In this case, we used the template
superhump light curve obtained in appendix \ref{sec:recon}.
Because the superhump profiles varied with time, we averaged
the template superhump profiles $\pm$1~d of the observations.
Since the superhump profile was not well determined for stage A,
we used the template for BJD 2455236 before this epoch.
For the same reason, we used the template for BJD 2455245
after this epoch.  The template was normalized to have full
amplitudes of 0.2 mag and the zero mean value.  The phase of
the maximum (determined by spline interpolation) was shifted
to phase zero.  The definition of the superhump maximum in
the present case refers to the light maximum, and is slightly
different from \citet{Pdot}, which more reflected the moment
of the variation.  This difference, however, was very subtle
and did not affect the overall $O-C$ appearance.

   We used observations in $\pm$1.5 superhump cycles around 
the superhump maximum to be measured.  These three superhump
cycles typically contains 9 observations.  We then used
a parameter set of \{time of maximum, amplitude\} to model
the observation.  The modeling was the same as in appendix 
\ref{sec:recon} and used numerical integration for 
the spline-interpolated template superhump light curve.
Using the likelihood function equation (\ref{equ:residual})
(no prior was used here), we can perform the MCMC calculation
to determine the PDF for the parameter set.  We adopted
the means and standard deviations of the MCMC results
(after removing the burn-in period) as the mean values
and 1$\sigma$ errors for the estimated time of superhump maximum
and amplitude.

\end{document}